\documentclass[twocolumn,superscriptaddress,amsmath,amssymb,aps,sort&compress,prl]{revtex4-2}
\usepackage[utf8]{inputenc}
\usepackage{times,color,amsthm,graphics,graphicx,bm,bbm,dcolumn}
\usepackage{epsfig}
\usepackage{graphicx}
\usepackage{siunitx}
\usepackage{xcolor}
\usepackage[colorlinks,urlcolor=blue,citecolor=blue,linkcolor=teal]{hyperref}

\setcitestyle{super}

\begin{document}

\title{Connecting shear-flow and vortex array instabilities in annular atomic superfluids}

\author{D. Hern\'andez-Rajkov}
\email[E-mail: ] {rajkov@lens.unifi.it}
\affiliation{European Laboratory for Nonlinear Spectroscopy (LENS), University of Florence, 50019 Sesto Fiorentino, Italy}
\affiliation{Istituto Nazionale di Ottica del Consiglio Nazionale delle Ricerche (CNR-INO) c/o LENS, 50019 Sesto Fiorentino, Italy}

\author{N.~Grani}
\affiliation{European Laboratory for Nonlinear Spectroscopy (LENS), University of Florence, 50019 Sesto Fiorentino, Italy}
\affiliation{Istituto Nazionale di Ottica del Consiglio Nazionale delle Ricerche (CNR-INO) c/o LENS, 50019 Sesto Fiorentino, Italy}
\affiliation{Department of Physics, University of Florence, 50019 Sesto Fiorentino, Italy}

\author{F.~Scazza}
\affiliation{Department of Physics, University of Trieste, 34127 Trieste, Italy}
\affiliation{European Laboratory for Nonlinear Spectroscopy (LENS), University of Florence, 50019 Sesto Fiorentino, Italy}
\affiliation{Istituto Nazionale di Ottica del Consiglio Nazionale delle Ricerche (CNR-INO) c/o LENS, 50019 Sesto Fiorentino, Italy}

\author{G.~Del~Pace}
\affiliation{Department of Physics, University of Florence, 50019 Sesto Fiorentino, Italy}

\author{W.~J.~Kwon}
\affiliation{Department of Physics, Ulsan National Institute of Science and Technology (UNIST), Ulsan 44919, Republic of Korea}

\author{M.~Inguscio}
\affiliation{European Laboratory for Nonlinear Spectroscopy (LENS), University of Florence, 50019 Sesto Fiorentino, Italy}
\affiliation{Istituto Nazionale di Ottica del Consiglio Nazionale delle Ricerche (CNR-INO) c/o LENS, 50019 Sesto Fiorentino, Italy}
\affiliation{Department of Engineering, Campus Bio-Medico University of Rome, 00128 Rome, Italy}

\author{K.~Xhani}
\affiliation{Istituto Nazionale di Ottica del Consiglio Nazionale delle Ricerche (CNR-INO) c/o LENS, 50019 Sesto Fiorentino, Italy}

\author{C.~Fort}
\affiliation{Istituto Nazionale di Ottica del Consiglio Nazionale delle Ricerche (CNR-INO) c/o LENS, 50019 Sesto Fiorentino, Italy}
\affiliation{Department of Physics, University of Florence, 50019 Sesto Fiorentino, Italy}

\author{M.~Modugno}
\affiliation{Department of Physics, University of the Basque Country UPV/EHU, 48080 Bilbao, Spain}
\affiliation{IKERBASQUE, Basque Foundation for Science, 48013 Bilbao, Spain}
\affiliation{EHU Quantum Center, University of the Basque Country UPV/EHU, 48940 Leioa, Biscay, Spain}

\author{F.~Marino}
\affiliation{Istituto Nazionale di Ottica del Consiglio Nazionale delle Ricerche (CNR-INO) c/o LENS, 50019 Sesto Fiorentino, Italy}
\affiliation{Istituto Nazionale di Fisica Nucleare, Sez.~di Firenze, 50019 Sesto Fiorentino, Italy}

\author{G.~Roati}
\affiliation{European Laboratory for Nonlinear Spectroscopy (LENS), University of Florence, 50019 Sesto Fiorentino, Italy}
\affiliation{Istituto Nazionale di Ottica del Consiglio Nazionale delle Ricerche (CNR-INO) c/o LENS, 50019 Sesto Fiorentino, Italy}

\begin{abstract}
At the interface between two fluid layers in relative motion, infinitesimal fluctuations can be exponentially amplified, inducing vorticity and the breakdown of the laminar flow. This process, known as the Kelvin-Helmholtz instability \cite{helmholtz, kelvin}, is responsible for many familiar phenomena observed in the atmosphere \cite{luce, fukao} and in the oceans \cite{li,smyth}, as well as in astrophysical objects \cite{McNally2012}, being known as one of the paradigmatic routes to turbulence in fluid mechanics \cite{klaassen, mashayek1, thorpe1, thorpe2}. While shear-flow instabilities in classical fluids have been extensively observed in various contexts, controlled experiments in the presence of quantized circulation are comparatively very few \cite{blaauwgeers, finne2006}. Here, we engineer two counter-rotating atomic superflows, a configuration that in classical inviscid fluids is unstable via the Kelvin-Helmholtz instability. We observe how the contact interface, i.e. the shear layer, develops into an ordered circular array of quantized vortices, which loses stability and rolls up into vortex clusters \cite{thorpe4}. We extract the instability growth rates and find that they obey the same scaling relations across different superfluid regimes, ranging from weakly-interacting bosonic to strongly-correlated fermionic pair condensates. The measured scalings, reproduced by numerical simulations and well described by a microscopic point-vortex model \cite{aref, havelock}, are consistent with the classical hydrodynamic Kelvin-Helmholtz instability of a finite-width shear layer \cite{rayleigh,drasin, charru}. Our results establish interesting connections between vortex arrays and shear-flow instabilities, suggesting a possible interpretation of the observed quantized vortex dynamics as a manifestation of the underlying unstable flow \cite{baggaley, carusotto}. Moreover, they open the way for exploring a wealth of out-of-equilibrium phenomena, from vortex-matter phase transitions \cite{Johnstone2019, Gauthier2019} to the spontaneous emergence and decay of two-dimensional quantum turbulence \cite{barenghi, kobyakov, Neely2013}.
\end{abstract}

\maketitle

A close relationship exists between vortices and shear-flow instabilities in fluid mechanics. In classical hydrodynamics, the interface between two fluid layers in relative motion is identified by an ideal surface containing an infinite number of line vortices, namely a vortex sheet \cite{charru}. More than one century ago, Lord Kelvin and von Helmholtz predicted the dynamical instability of such a vortex sheet \cite{helmholtz, kelvin}, later generalized by Rayleigh to the case of a finite-width shear layer \cite{rayleigh}. The Kelvin-Helmholtz instability (KHI) initially manifests itself as a wave-like deformation of the interface, exponentially growing in time with a rate proportional to the relative velocity between the two fluids. It quickly leads to the twisting of the vortex sheet \cite{drasin,charru} and eventually to a turbulent mixing of spiraling structures \cite{thorpe2}. Starting from the seminal experiments by Reynolds in 1883 \cite{reynolds}, the KHI has been the subject of extensive research and experimentation \cite{thorpe, kent, shearer}.

The last few decades have seen exciting advances in the field of superfluidity  -- in particular in ultracold atomic systems \cite{novelSF2} -- owing to an ever-increasing capability to manipulate pristine quantum systems in which vorticity is quantized, and dissipation occurs through channels different from those in ordinary fluids. A natural question is whether these key differences affect the onset and the microscopic nature of flow instabilities and to what extent classical scaling relations apply to superfluids, particularly in the presence of strong interactions. Theoretical investigations in quantum fluids have been mostly focused on the stability of the interface between distinct sliding fluid components, such as superfluid and normal phases of the same liquid \cite{korshunov, Volovik2002, Korshunov2002},and in binary Bose-Einstein condensates \cite{takeuchi, suzuki, lundh, Kokubo2021} (BECs). Experimental observations are limited to the interface between the A and B phases of ${}^3$He in a rotating cryostat \cite{blaauwgeers, finne2006}. More recently, the appearance of streets of quantized vortices in rapidly rotating atomic BECs has been associated with Kelvin-Helmholtz dynamics\cite{zwierlein}.

\begin{figure*}[t!]
\centering
\vspace{0 pt}
\includegraphics[width=1.35\columnwidth]{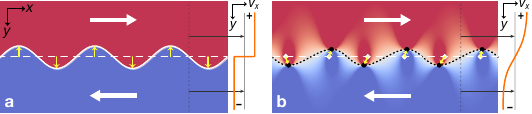}
\caption{
\textbf{Instability of classical and quantum counter-propagating shear flows.} 
\textbf{a}, The (zero-width) vortex sheet separating two counter-propagating classical flows at time $t=0$ (horizontal dashed line) is unstable due to the KHI, developing at $t>0$ a wave-like deformation (solid line) with characteristic wave vectors that grow exponentially in time. 
\textbf{b}, In single-component superfluids, a shear-flow instability manifests itself as the deformation of the discrete vortex sheet \cite{baggaley, carusotto} formed by an array of quantized vortices (dots). Except at the position of the vortices, the tangential velocity gradually changes from the bulk value of one region to that of the other, giving rise to an effective finite-width shear layer.. Similarly to a real interface separating different sliding fluids, this layer (dotted line) is unstable and breaks down as evidenced by the vortex motion \cite{aref, havelock}. In both panels, background colors display the magnitude of the velocity component tangential to the initial shear layer. Vertical cuts of the tangential velocity profiles along the transverse direction are also depicted on the right of each panel (orange lines).} 
\label{fig:Fig1}
\end{figure*}

In general, the only necessary condition for the onset of an instability is a shear flow \cite{charru} -- i.e.,~two adjacent layers flowing at different velocities within a single homogeneous fluid (see Fig.~\ref{fig:Fig1}a). Here, we realize such a minimal scenario by engineering two counter-propagating flows in a single-component atomic superfluid. Recently, this approach has been proposed as a possible route to KHI in quantum fluids and numerically validated for atomic BECs~\cite{baggaley, carusotto}. Owing to the continuity of the wave function and the quantization of circulation, superfluids cannot support a continuous vortex sheet. Instead, a regular array of quantized vortices forms along the shear layer \cite{baggaley, carusotto} acting as a discrete version of a vortex sheet (see Fig.~\ref{fig:Fig1}b). The vortex array is unstable, and its dynamics can be directly mapped onto the instability of the associated counter-propagating flows, establishing a direct relationship between shear flow and vortex rows instabilities \cite{aref, havelock}.
In our experiment, we create a circular shear layer by exciting two counter-rotating superflows in a ring-shaped geometry. We observe the formation of a periodic, circular array of quantized vortices -- a vortex \emph{necklace} -- which rapidly breaks down, with nearby vortices rolling up and eventually displaying complex correlated trajectories. We demonstrate that the departure from the circular vortex necklace proceeds according to characteristic rates described by universal scaling relations across all the different superfluid regimes. Interestingly, the measured scalings are compatible with the classical KHI of a finite-width shear layer. Our observations establish atomic Fermi superfluids as a versatile laboratory for quantum fluid-dynamics experiments. We overcome typical difficulties that affect experiments with helium quantum fluids \cite{blaauwgeers,finne2006}, especially in terms of single-vortex detection and reconstruction of vortex trajectories \cite{Kwon}.

\begin{figure*}[t!]
\centering
\vspace{0 pt}
\includegraphics[width=\textwidth]{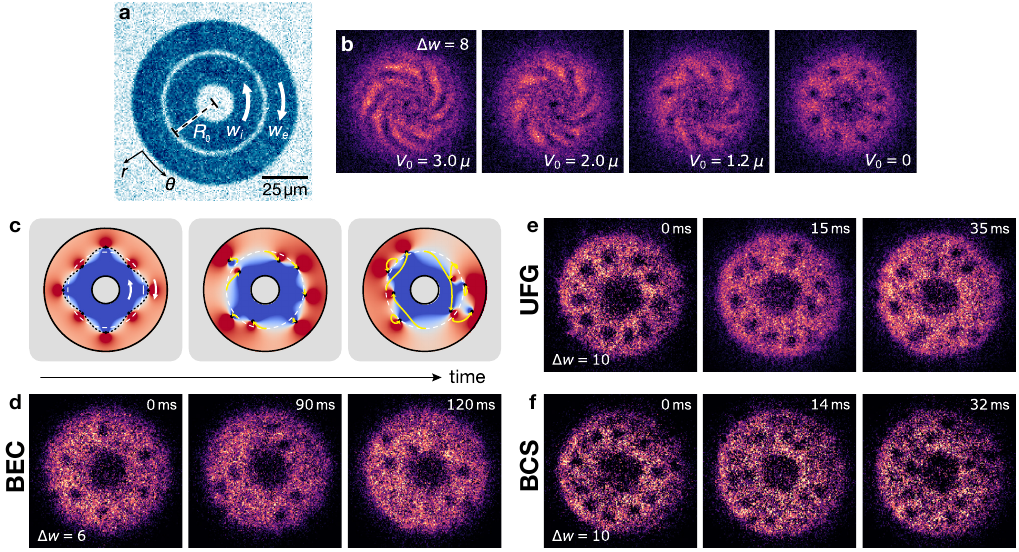}
\caption{\textbf{Shear-flow preparation and and emerging vortex dynamics.} 
\textbf{a}, \textit{In situ} density profile of two concentric counter-rotating superfluids, set in relative motion by a phase-imprinting technique while being separated by a potential barrier with FWHM $=\SI{1.4(1)}{\mu m}$. 
\textbf{b}, Single-shot TOF images in the BEC regime for $\Delta w=8$, as the barrier height $V_0$ is gradually lowered from $V_0\approx 3.5\,\mu$ to 0, where $\mu$ is the superfluid chemical potential (Methods). A necklace of $N_v=8$ vortices spontaneously forms at the location of the shear layer. 
\textbf{c}, PVM simulations of the dynamics of a necklace with $N_v=8$ vortices. In the first panel, the dotted sinusoidal line illustrates the interface mode $m=4$. Vortex trajectories are indicated by the yellow curved arrows while the background colors refer to the magnitude of the tangential component of the velocity, with the flow directions specified by white arrows. 
\textbf{d--f}, Typical single-shot TOF images of the vortex patterns obtained at different times $t\geq0$ after merging the two superfluids in distinct interaction regimes: \textbf{d.} BEC at $1/k_Fa=4.3(1)$ with $\Delta w=6$; \textbf{e}, UFG at $1/k_Fa=0.0(1)$ with $\Delta w=10$; and 
\textbf{f}, BCS superfluid at $1/k_Fa=-0.3(1)$ with $\Delta w=10$. In all interaction regimes, the vortex necklace destabilizes for $t>0$, and few-vortex clusters form.}
\label{fig:Fig2}
\end{figure*}

Our experiment starts with two thin and uniform Fermi superfluids comprising $N_p \simeq 3\times 10^4$ pairs of fermionic $^6$Li atoms (see Methods). Interactions between atoms forming the pairs are encoded in the $s$-wave scattering length $a$. This can be tuned through a broad Feshbach resonance, entering different superfluid regimes ranging from weakly interacting BECs of tightly bound molecules ($1/k_Fa>1$) to strongly correlated unitary Fermi gases (UFG, $1/k_Fa\simeq0$) and Bardeen-Cooper-Schrieffer (BCS, $1/k_Fa < 0$) superfluids. Here, $k_F = (ME_F/\hbar^2)^{1/2}$ is the Fermi wavevector, estimated from the system global Fermi energy $E_F$, and $M$ is the mass of a fermion pair. The superfluids are confined into concentric annular optical traps initially separated by a narrow potential barrier, resulting in two reservoirs with equal density $n_{2D}=\SI{4.96(2)}{\mu m^{-2}}$ (see Fig.~\ref{fig:Fig2}a). Sample temperatures are well below the critical temperature $T_c$ of the superfluid transition, $T=0.3(1)\,T_c$, corresponding to near-unity superfluid fractions in all interaction regimes. The superfluid healing length $\xi$ is always smaller than the vertical cloud size, making the superfluid dynamics three-dimensional. In contrast, vortex dynamics remain two-dimensional since only a few Kelvin modes of vortex lines are accessible \cite{supp}. 

We excite persistent flows in each reservoir by optically imprinting a dynamical phase onto the superfluid rings \cite{delpace}. In particular, we drive independent currents by shining oppositely oriented, azimuthal light gradients onto the internal and external rings (see Methods). With this procedure, we set integer quantized circulations $w_{i,e}$ with corresponding velocity fields $v_{i,e}(r)=\hbar w_{i,e}/(M r)$, where $r$ is the distance from the center, and the indices $i,e$ refer to the internal and external rings, respectively. The relative velocity at the interface equals $\Delta v = \hbar (w_e-w_i)/ (M R_0)$, where $R_0$ is the radius of the circular potential barrier separating the reservoirs. In all interaction regimes, we fix $w_e = - w_i$ and only consider velocities $\Delta v_\mathrm{max} < 0.7\,c_s$, where $c_s$ is the measured speed of sound in the bulk \cite{supp}.

We merge the two counter-rotating superfluids by gradually lowering the barrier potential. To follow the evolution of the flow, we image the atomic density profile after a short time-of-flight (TOF) expansion at varying times during the barrier removal. As shown in Fig.~\ref{fig:Fig2}b, as long as the barrier separates the two superfluid rings, a spiral interference pattern is observed due to the presence of an azimuthal phase gradient between the rings. The number of spiral arms matches the number $\Delta w \equiv|w_e-w_i|$ of $2\pi$ phase-slips at the circular interface \cite{eckel,delpace}. Eventually, when the barrier is completely removed, and the two superfluids come into full contact ($t=0$), we observe a necklace of $N_v\simeq\Delta w$ singly-charged, quantized vortices \cite{baggaley,carusotto} with the same circulation sign determined by vorticity (see Methods). Soon after the vortex necklace emerges, its periodicity breaks down, and vortices start pairing up in a quasi-synchronous process. As time progresses, metastable clusters of increasingly larger size form, following trajectories reminiscent of the characteristic Kelvin-Helmholtz roll-up dynamics (Fig.~\ref{fig:Fig2}c-f).

\begin{figure*}[ht!]
\centering
\vspace{0 pt}
\includegraphics[width=\textwidth]{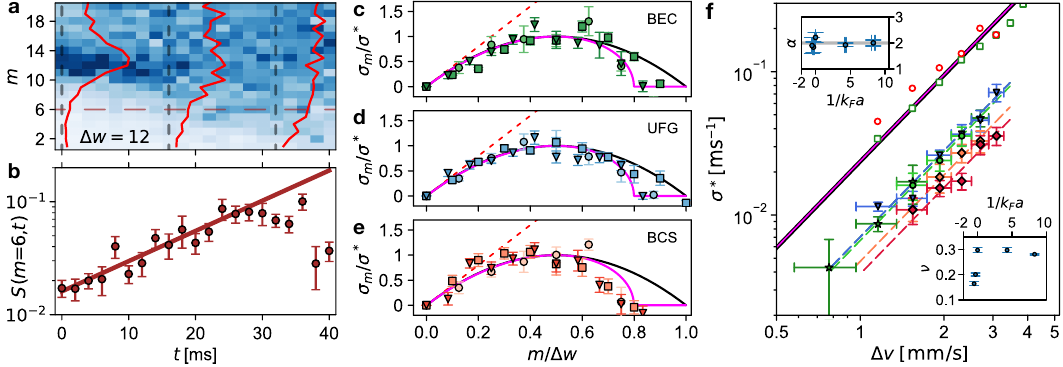}\vspace{-5 pt}
\caption{\textbf{Analysis of the vortex-array instability across the BCS-BEC crossover.}
\textbf{a}, Normalized angular structure factor $s(m, t)$ for a UFG superfluid and $\Delta w=12$. $s(m, t)$ is extracted by averaging over $\sim 20$ experimental realizations. Solid red lines represent vertical cuts $s(m)$, corresponding to angular spectra at different times, and the dashed horizontal line indicates the expected most unstable mode. 
\textbf{b}, Time evolution of $s$ for the most unstable mode, $m^*=6$. An exponential fit of the data (line) is used to extract the growth rate $\sigma^*$. Error bars denote the standard error of the mean over $\sim 20$ experimental realizations.
\textbf{c--e}, Normalized dispersion relations, $\sigma_m/\sigma^*$, as a function of $m/\Delta w$ for: \textbf{c}, BEC [$1/k_Fa=4.3(1)$]; \textbf{d}, UFG [$1/k_Fa=0.0(1)$]; and \textbf{e}, BCS [$1/k_Fa=-0.3(1)$] superfluids. Rates are shown for different $\Delta w=8,10,12$, respectively denoted by circles, squares, and triangles. The red dotted line shows the low-wavenumber limit, $\frac{1}{2}k \Delta v$ normalized to $\sigma_\mathrm{PVM}^*$, while solid lines show the rates predicted by the PVM in Eq.~(\ref{eq:pvm}) (black line) and by Rayleigh's Eq.~(\ref{eq:RayleighFormula}) using $\delta=0.8\hbar/M\Delta v$ (magenta line) (see Ref.~\citenum{supp} for the determination of $\delta$). 
\textbf{f}, Scaling of $\sigma^*$ against $\Delta v$ in different interaction regimes. Filled symbols correspond to BCS ($1/k_Fa=-0.5(1)$, red diamonds), ($1/k_Fa=-0.3(1)$, orange diamonds); UFG ($1/k_Fa=0.0(1)$, blue triangles); and BEC ($1/k_Fa=4.3(1)$, dark green stars), ($1/k_Fa=8.3(3)$, light green circles). Open symbols refer to GP simulations at $T=0$ (open green squares) and cZNG simulations at $T/Tc=0.4$ (open red circles) for $1/k_Fa=4.3(1)$. Error bars are smaller than the symbols. Solid lines refer to  Rayleigh's (magenta) and PVM's (black) predictions. Dashed lines denote fits with $\sigma^* = A{\Delta v}^{\alpha}$. As insets, as a function of $1/k_Fa$: (top) fitted scaling exponents $\alpha$; (bottom) adimensional factor $\nu$ defined as $\sigma^* = \nu \,\sigma_\mathrm{PVM}^*$. In panels, \textbf{c}--\textbf{f}, vertical error bars denote fitting $1\sigma$-errors, while horizontal ones reflect the experimental uncertainty on the initial vortex number.
}
\label{fig:Fig3}
\end{figure*}

We reconstruct the evolution of the vortex necklace by characterizing it through the vortex structure factor\cite{Warren}. In our circular geometry, although vortices move both in radial and azimuthal directions, the vortex array modes are mainly azimuthal (see dotted line in Fig.~\ref{fig:Fig2}c). To decompose the motion, it is natural to introduce the angular structure factor, defined as $S(m, t)=(1/N_v)\sum_{j,l} \exp[i m(\theta_j(t)-\theta_l(t))]$. Here, $\theta_j(t)$ is the angular position of the $j$-th vortex, and $m$ is the integer winding number of the mode, defined as $m = k R_0$, where $k$ is the mode wavenumber. Figure \ref{fig:Fig3}a displays the average $s(m, t)=S(m, t)/N_v$ measured for $1/k_Fa=0.0(1)$ and $\Delta w=12$. The spectral peak at $m=\Delta w$, characteristic of a periodic necklace, evolves towards lower angular modes while simultaneously broadening. Similar behavior is found in all interaction regimes. The increasing population of lower modes identifies the breakdown of the necklace structure. For most modes $m$, we observe that $s(m,t)$ grows exponentially in time as $s(m,t) \sim e^{2 \sigma_m t}$, where $\sigma_m$ is the growth rate of the $m$-th mode. As an example, in Fig.~\ref{fig:Fig3}b, we present the behavior for $m=6$.

We analyze the spectral dependence of $\sigma_m$ for different relative velocities using the point vortex model \cite{supp} (PVM). This model treats each vortex as a point particle advected by the velocity field generated by all other vortices. The vortex motion is thus determined by the background flow, which, in turn, emerges as a phenomenon associated with the vortex dynamics. As discussed in \cite{aref, havelock}, an instability appears as the departure from the initial ordered vortex configuration with characteristic rates given by

\begin{equation}\label{eq:pvm}
\sigma_\mathrm{PVM}(k, \Delta v) = \frac{\Gamma k}{2d_v}\left(1-\frac{kd_v}{2\pi}\right),
\end{equation}

\noindent where $\Gamma=h/M$ is the quantized vortex circulation, and $d_v=h/(M\Delta v)$ is the initial inter-vortex separation. The maximum growth rate $\sigma^*$, i.e.~the growth rate of the most unstable mode, is reached for $k^*=\pi/d_v$, in agreement with the numerical results reported in Ref.~\citenum{carusotto}, where the instability of a linear array of vortices in a single-component BEC is characterized by solving the corresponding Bogoliubov problem.
This particular wavenumber sets a fundamental scaling law, $\sigma^*\propto {\Delta v}^2$, providing an experimentally accessible hallmark of the instability. In the low-wavenumber limit, $kd_v \ll 1$, Eq.~(\ref{eq:pvm}) reduces to $\sigma_\mathrm{PVM} \sim \frac{1}{2}k \Delta v$, which equals Kelvin's growth rate of a continuous vortex sheet\cite{charru} (i.e., a zero-width interface). Hence, the PVM extends the classical KHI scenario to the case of a \textit{discrete} vortex sheet formed by a finite number of vortices. It also predicts a smooth change of the tangential flow between the two layers, except at the position of the vortices, resulting in a finite-size shear layer with half-width $\delta = \hbar/(M \Delta v)$ (Fig.~\ref{fig:Fig1}b, and Ref. \citenum{supp}). 

In our regime of shear velocities, the instability rates have been found to follow the Rayleigh model for the KHI of a finite-width classical shear layer~\cite{carusotto}. Although here we extract the rates from the vortex trajectories and not directly from Bogoliubov excitations, we find it interesting to compare our data with the predictions of this classical model \cite{supp}.

In Fig.~\ref{fig:Fig3}c-e, we show the normalized dispersion relations $\sigma_m/\sigma^*$ as a function of $m/\Delta w$ measured for BEC, UFG, and BCS superfluids. All the normalized rates are in good agreement with the expected trends from the PVM [Eq.~(\ref{eq:pvm})] and Rayleigh [Eq.~(\ref{eq:RayleighFormula}), Methods] formulas. The agreement is more evident for $k \lesssim k^*$, where the rates predicted by the two models are nearly identical, suggesting a close relationship between the instability of the vortex necklace and that of the associated shear-flow \cite{carusotto}. In Fig.~\ref{fig:Fig3}f, the extracted $\sigma^*$ for different superfluid regimes is plotted as a function of the relative velocity, displaying the expected quadratic behavior.
These scaling properties, together with the normalized dispersion relations in Fig.~\ref{fig:Fig3}c-e. suggest an interpretation of the dynamics as a quantized analogue of the Kelvin-Helmholtz scenario across the BEC-BCS crossover.

The data in Fig.~\ref{fig:Fig3}f are compared with Eqs.~(\ref{eq:pvm}) and (\ref{eq:RayleighFormula}), and with 3D numerical simulations based on Gross-Pitaevskii (GP) equation and collisionless Zaremba-Nikuni-Griffin (cZNG) model\cite{ZNGbook}. While theoretical rates -- analytical and numerical -- agree quantitatively with each other, the measured ones show systematically lower values, quantified by the ratio $\nu=\sigma^*/\sigma_\mathrm{PVM}^*$. In classical fluids, dissipation effects (e.g., surface tension, viscosity) typically stabilize the system, leading to lower growth rates \cite{charru, drasin, Villermaux1998,Betchov1963}. In our system, we expect finite temperature to introduce additional dissipative effects through the mutual friction between the superfluid and the non-vanishing normal components. The microscopic origin of this source of dissipation resides in scattering processes between normal excitations and vortices, hence strongly depend on the vortex core structure\cite{Kopnin2002, Sonin2015}. For instance, in the fermionic regimes, the presence of Andreev quasiparticles localized in the vortex core give rise to additional dissipation channels \cite{Silaev2012, Sonin2015} that should  become less relevant moving towards the BEC regime\cite{barresi}. These effects can be phenomenologically included in the PVM, and modify the instability rates accordingly (see Methods). In an attempt to take into account temperature effects, we have performed numerical simulations of the dynamics using the cZNG model, which includes thermal-condensate interactions only at a mean-field level. As shown in Fig.~\ref{fig:Fig3}f, the obtained rates are comparable with those from $T=0$ GP simulations. This approach does not account for the slower rates observed in the experiment. A more complete\cite{ZNGbook, Allen2013} or alternate microscopic model\cite{mehdi, Sergeev2023} should be used to include further effects of the normal component on vortex dynamics. We remark that a quantitative understanding of the microscopic mechanisms connecting non-local vortex dynamics to dissipation, especially in the presence of strong correlations, remains an open problem both experimentally and theoretically \cite{Kopnin2002, Sonin2015, mehdi}. In addition, we cannot exclude that technical imperfections such as spurious excitations originating from the barrier removal and a non-ideal imprinting procedure (not considered in any of the performed numerical simulations), \cite{delpace, supp} may potentially foster additional dissipation. Finally, we would like to stress that -- essentially due only to quantization of the vorticity in units of $\Gamma$ -- the maximum growth rate of the instability takes the universal form $\sigma^* \propto (\Delta v)^2/\Gamma$, irrespective of the considered pairing regime across the BCS-BEC crossover. Interestingly, the proportionality constant, i.e. $\pi /4 \times \nu$, is of order $1$ in all regimes.

\begin{figure}[t!]
\centering
\includegraphics[width=1\columnwidth]{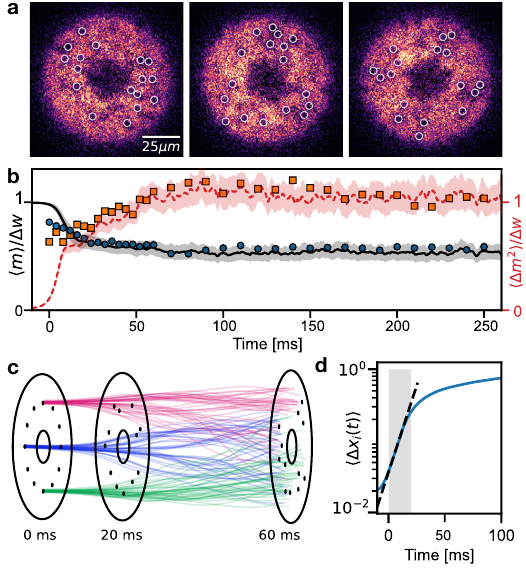}\vspace{-5pt}
\caption{\textbf{Nonlinear vortex dynamics.} 
\textbf{a}, Single-shot TOF images of vortex patterns acquired after $36$\,ms of evolution, showing $2-$, $3-$, and $4-$fold vortex cluster symmetries. Images are acquired for a BEC superfluid starting from nominally identical necklaces with $N=16$. b
\textbf{b}, Time evolution of the mean winding number $\langle m \rangle$ (circles) and its variance $\langle \Delta m^2 \rangle$ (squares) normalized to $\Delta w$ for $1/k_Fa=4.3(1)$ and $\Delta w=12$, averaged over $\sim\,$20 experimental realizations. The data are compared with PVM simulations (black-solid and red-dashed lines). 
\textbf{c}, Trajectories of three vortices (colored lines) belonging to nearly-identical necklaces with $N=12$. A set of 40 initial vortex positions are picked randomly, each within a range of one healing length $\xi \approx 0.5\mu$m around their reference value, for which the necklace is perfectly periodic. We show the vortex pattern at $t=0,20,60$ms for one realization. 
\textbf{d}, Mean-distance between each-vortex trajectory calculated as a function of time by averaging over $40$ nearly-identical random initial conditions, as described above, and over $N_v=12$ vortices. The distances are normalized to the mean separation of any given two points in the ring geometry (Methods). At short times, $t<20$ ms (shaded area), the separation between trajectories grows exponentially (dashed line) with a characteristic maximal Lyapunov exponent of $\Lambda=0.111(5)\,\text{ms}^{-1}$, in agreement with the maximum instability growth rate $\sigma_\mathrm{PVM}^*=0.127(5)$ obtained for $\Delta w=12$. Shaded regions around lines denote the standard deviation of the mean.
}
\label{fig:Fig4}
\end{figure}

After the initial stage of the instability, with characteristic time $1/\sigma^*$, the vortex dynamics enter a nonlinear regime with vortex clusters forming and fragmenting as time progresses while conserving the total vortex number. Starting from nominally identical initial conditions, vortices form clusters with different symmetries, as shown in Fig.~\ref{fig:Fig4}a. Which of these symmetries appears at a given time hinges on the initial conditions and fluctuations in the system. Averaging over different realizations, we observe the system exploring widely different configurations, losing information about the initial order. In particular, the mean winding number $\langle m \rangle$ is nearly constant in time around $\Delta w/2$, while the variance $\langle \Delta m^2 \rangle$ saturates to the vortex number $\Delta w$ (see Fig.~\ref{fig:Fig4}b). This indicates that the system reaches a steady state, where all modes are significantly populated. PVM simulations starting with nearly identical configurations reproduce the experimental data quantitatively, suggesting that vortices tend to spread over the whole system volume (Fig.~\ref{fig:Fig4}c).

Signals of the instability are encoded even at the single-vortex level. We compare the vortex trajectories of the different PVM simulations starting from nearby initial conditions and compute their relative separation as a function of time (see Fig.~\ref{fig:Fig4}c-d and Methods). The average separation between same-vortex trajectories grows exponentially, $\sim e^{\Lambda t}$, with a rate $\Lambda$ similar to the maximal growth rate, $\sigma_\mathrm{PVM}^*$ (Fig.~\ref{fig:Fig4}d). While $\sigma^*$ quantifies the collective motion of the necklace, the divergence rate $\Lambda$ -- known as the maximal Lyapunov exponent \cite{pikovsky2016} -- refers to the trajectory of a single vortex. A positive $\Lambda$ is here indicative of divergent vortex trajectories starting from arbitrarily close initial conditions of the necklace, and not of chaotic dynamics. This connection clarifies the role of quantized vortices in defining a shear layer and the associated instability as the mechanism behind the breakdown of the vortex necklace. For $t \gg 1/ \Lambda$, the distance between trajectories tends to saturate to the average separation of two random points in the system (set by the value $1$ in the graph). This is due to the finite-size effects that constrain the trajectories divergence, eventually leading to the attainment of boundary-dominated dynamical equilibrium.  On the other hand, such equilibrium hides complex underlying dynamics. Chaotic advection and turbulent flows are expected \cite{babiano} for point-vortex systems for $N_v>2$. Classically, shear-flow instabilities such as the KHI in inviscid fluids drive the system into a turbulent state portrayed by an irregular -- sensitive to initial conditions -- mixing of spiraling structures at different scales \cite{thorpe2}. The intertwined vortex trajectories shown in Fig.~\ref{fig:Fig4}c are reminiscent of this scenario.

%--------------------------------------------------
%------------------ Conclusions -------------------
%--------------------------------------------------
Our observation of a shear-flow instability in atomic superfluids showcases a pristine example of an emergent phenomenon, with quantized vortices acting simultaneously as sources and probes of the unstable flow. The same microscopic mechanism operates in all superfluid regimes, and it underlies the observed common behavior, which belongs to the class of classical inviscid fluids with finite-size shear layers. Our work ushers in the exploration of the fundamental connections between quantized vortex dynamics in scalar single-component superfluids and classical shear-flow instabilities, posing interesting questions on how the limit of a continuous vortex sheet may be reached within a discretized point-vortex scenario. We anticipate our results to be of relevance for diverse non-equilibrium phenomena in strongly correlated quantum matter, ranging from rapidly rotating quantum gases \cite{zwierlein} to pulsar glitches \cite{Haskell2015} and neutron star mergers \cite{Price2006}. Our findings also set the starting point to explore a variety of vortex matter phase-transitions in fermionic superfluids \cite{Sachkou2019},  including negative temperature and non-trivial cluster states \cite{Simula, Johnstone2019, Gauthier2019, Reeves2022}, even in presence of dissipative mechanisms resulting from vortex-vortex and vortex-quasiparticles interactions \cite{Heyl,mehdi}. An exciting direction for future experiments concerns the cascade of secondary instabilities towards the spontaneous onset of quantum turbulence \cite{baggaley, kobyakov, finne2006}, exploring a route complementary to external forcing \cite{Henn2009, Navon2016} to probe its underlying microscopic mechanisms from the few- to the many-vortex perspective.

\vspace*{-10pt}
\section{Acknowledgements}
We thank Iacopo Carusotto, Nigel Cooper, and Giovanni Modugno for their valuable comments on the manuscript and the Quantum Gases group at LENS for fruitful discussions. This work was supported by the European Research Council (ERC) under Grant Agreement No.~307032, the Italian Ministry of University and Research under the PRIN2017 project CEnTraL and PNRR project PE0000023-NQSTI, the European Union’s Horizon 2020 research and innovation program under the Qombs project FET Flagship on Quantum Technologies Grant Agreement No.~820419. W.~J.~K acknowledges support from the Research Fund (1.220137.01) of UNIST (Ulsan National Institute of Science and Technology). M.M.~acknowledges support from Grant No.~PID2021-126273NB-I00 funded by MCIN/AEI/10.13039/501100011033 and ``ERDF - A way of making Europe'', and from the Basque Government through Grant No.~IT1470-22. F.~S.~acknowledges funding from the European Research Council (ERC) under the European Union’s Horizon 2020 research and innovation programme (Grant agreement No.~949438) and from the Italian MUR under the FARE programme (project FastOrbit).

%--------------------------------------------------
%--------------------- Methods --------------------
%--------------------------------------------------

\section{Methods}

\subsection{Sample preparation}
We prepare fermionic superfluid samples by evaporating a balanced mixture of the two lowest hyperfine spin states $|F,m_F\rangle= |1/2,\pm1/2\rangle$ of $^6$Li, near their scattering Feshbach resonance at $\SI{832}{G}$ in an elongated, elliptic optical dipole trap, formed by horizontally crossing two infrared beams at a $14^{\circ}$ angle. At the end of the evaporation, we sweep the magnetic field to the desired interaction regime. A repulsive $\mathrm{TEM_{01}}$-like optical potential at $\SI{532}{nm}$ with a short waist of about $\SI{13}{\mu m}$ is then adiabatically ramped up before the end of the evaporation to provide strong vertical confinement, $\omega_z\simeq 2\pi\times400$ Hz. Successively, a box-like potential is turned on to trap the resulting sample in a circular region of the $x$–$y$ plane. This circular box is tailored using a Digital Micromirror Device (DMD). When both potentials have reached their final configuration, the infrared lasers forming the crossed dipole trap are adiabatically extinguished, completing the transfer into the final uniform pancake trap\cite{Kwon}. Finally, to create the pair of superfluid rings at rest, we dynamically change the DMD-tailored potential. We first create the hole at the center of the initial disk and then dynamically increase its size until reaching a radius of $R_i=\SI{10.0\pm 0.2}{\mu m}$. Finally, an optical barrier separating the two superfluid rings is adiabatically raised at $R_0=\SI{27.5\pm 0.2}{\mu m}$. A residual radial harmonic potential of $\SI{2.5}{Hz}$ is present due to the combined effect of an anti-confinement provided by the $\mathrm{TEM_{01}}$ laser beam in the horizontal plane and the confining curvature of the magnetic field used to tune the Feshbach resonance. This weak confinement has a negligible effect on the sample over the $R_e=\SI{45.0\pm 0.2}{\mu m}$ radius of our box trap, resulting in an essentially homogeneous density.

\vspace{-10pt}
\subsection{Phase-imprinting procedure}
We excite controllable persistent current states in each of the two rings by using the phase imprinting protocol described in Ref.~\citenum{delpace}. Using the DMD, we create an optical gradient along the azimuthal direction, namely $U(r, \theta) = U_0 \,\theta/2\pi \times \mathrm{sign}(r-R_0)$. By projecting such a potential over a time $t_I<\hbar /\mu$, we imprint a phase $\phi(r,\theta)= U(r,\theta) \,t_I/h$ onto the superfluid wavefunction. By suitably tuning the imprinting time $t_I$ and the gradient intensity $U_0$, we excite well-defined winding number states in each of the two rings in a reproducible way. We measure the imprinted circulations using an interferometric probe: we let the two rings expand for $3$ ms of time of flight (TOF) and then image the resulting spiral-shaped interference pattern. The first panel of Fig. \ref{fig:Fig2}b shows an interferogram for $\Delta w = 8$. In particular, the number of spirals in the interferogram yields the relative winding number $\Delta w$ between the two rings \cite{eckel,delpace}. Additionally, we independently check that before the imprinting procedure, the inner ring is in the $w=0$ state by realizing a similar experimental protocol now in a geometry similar to the reported in Ref. \citenum{delpace}. All circulation states excited in the two rings have been observed to persist for several hundreds of ms \cite{delpace}, except for $\Delta w > \mathrm{12}$ in the BCS regimes. Nevertheless, we observe these states to not decay for the typical timescale of the observed instability $t <\SI{40}{ms}$. To reduce the effect of extra density excitation on the dynamics, we wait $\SI{300}{ms}$ after imprinting before removing the barrier between the superfluids.

\vspace{-10pt}
\subsection{Vortex imaging and tracking}
We establish a shear flow by removing the circular barrier separating the two ring superfluids. In particular, we lower down its intensity by opportunely changing the DMD pattern. The barrier removal process takes 28 ms and brings the system into the vortex necklace configuration of Fig \ref{fig:Fig2}. We confirm that the duration of the barrier removal does not significantly affect the dynamics. Removing the barrier over time scales faster than 10\,ms creates unwanted excitations such as solitonic structures. To image the vortices in the BEC regime, we acquire the TOF image of the superfluid density, where vortices appear as clear holes. In particular, we abruptly switch off the vertical confinement and at the same time, we start to ramp down the DMD potential, removing it completely in $\SI{1}{ms}$. Then, we let the system evolve further for $\SI{2.2}{ms}$ of TOF and then acquire the absorption image. This modified TOF method allows for to maximization of the vortex visibility. However, the small condensed fraction in the strongly-interacting regime makes it impossible to detect vortices with this simple method. Therefore, in the UFG and BCS regimes, we employ the technique developed in Ref. \citenum{Kwon}: we add a linear magnetic field ramp of $4-5$ ms to $\SI{700}{G}$ before the imaging to map the system in a BEC superfluid. The position of the vortices is tracked manually in each acquired image. The size of the vortex limits the error on the position of the vortex after the TOF sequence. To estimate it, we perform a Gaussian fit of the vortex density hole and obtain a waist of $\sim 1.0-\SI{1.4}{\mu m}$ for all interaction regimes.

\subsection{Quasi-2D vortex dynamics}
The vertical confinement provided by the $\mathrm{TEM_{1,0}}$ laser beam is such that in the BEC regimes, the ratio $\mu/(\hbar \omega_z) \gtrapprox 1.5$, and in the UFG and BCS regimes $E_F/(\hbar \omega_z) \approx 6$; making the system collisionally three dimensional. However, vortex dynamics behave as a quasi-two-dimensional system since only a few Kelvin modes can be populated. In fact, the standard Kelvin dispersion \cite{Rooney2011} is:
\begin{equation}
    \omega(k) = -\frac{\hbar k^2}{2M} \log\left(\xi k\right),
\end{equation}
where $\xi$ is the healing length. Due to geometrical restrictions \cite{Rooney2011}, only modes with wavelength larger than the healing length can be effectively populated in a superfluid. Under our experimental condition, this translates into the fact that only the lowest wavenumber Kelvin mode with $k=\pi/R_z$ can be populated in the BEC regime, where $R_z=\sqrt{2\mu/(M\omega_z^2)}$ is the Thomas-Fermi radius in the $z$-direction. On the other hand, in the UFG and BCS regimes, due to higher Thomas-Fermi radius and smaller healing length, only the first three Kelvin modes with $k_n=(\pi+2\pi n)/R_z$ can be populated. Anyway, in all the interaction regimes explored in this work, the number of possibly populated Kelvin modes remains so small that we can assume a 2D dynamics of the vortex motion.

\subsection{Preparation of the vortex necklace}
We remove the circular barrier (Fig.~\ref{fig:Fig2}a) between the two rings by lowering its intensity using a sequence of 15 different DMD patterns. To obtain a clear initial condition of the vortex crystal and to prevent the formation of other excitation in the system \cite{kanai2019merging}, we set the duration of the barrier removal to $\tau = \SI{28}{ms}$.

After the complete barrier removal, we observe the creation of a vortex necklace with a number of vortices given by the relative circulation $\Delta w$, as illustrated in Fig.~2a. The phase imprinting method allows to excite circulation states in the two superfluids in a highly reproducible way, but experimental imperfections can lead to shot-to-shot fluctuations in the circulation state of the rings. This leads to fluctuations in the initial configuration of vortices, which we estimate by analyzing the statistics of the relative circulation and vortex number in datasets of 100 experimental realizations on a BEC superfluid at $1/k_Fa_s=4.1(1)$. Figure~\ref{fig:ExpMeth2}a shows the distribution of the measured relative circulation between the two rings $\langle \Delta w\rangle_M$ with respect to the target $\Delta w_T$, measured from interferograms acquired before removing the circular barrier for $\Delta w_T = 6$ and $\Delta w_T=12$. In Fig.~\ref{fig:ExpMeth2}b, the number of spurious vortices introduced by the phase-imprinting protocol is displayed, measured from the TOF expansion of the two rings before the barrier removal. Finally, Fig.~\ref{fig:ExpMeth2}c shows the distribution of the total number of vortices in the superfluid detected in the TOF expansion after removing the circular barrier. 
Despite the high reliability in producing the desired circulation states in the two rings(Fig.~\ref{fig:ExpMeth2}a), we observe that the distribution of the total number of vortices detected after the barrier removal is augmented and broadened by the presence of  spurious vortices. This leads to residual fluctuations of the initial configurations of the vortex necklace (Fig.~\ref{fig:ExpMeth2}d), which determine the experimental uncertainty on the initial relative velocity $\Delta v$ (see horizontal error bars in Figs.~3f,g). They also contribute to the experimental noise on the extracted exponential growth rate for a given $\Delta w$.

\subsection{Rayleigh model}
In classical fluid mechanics, the problem of the stability of a finite-width shear layer was first analyzed by Rayleigh \cite{rayleigh}, who derived an interface-dependent growth rate as:
\begin{equation}\label{eq:RayleighFormula}
    \sigma_\mathrm{R}(k, \Delta v) = \mathrm{Im} \frac{\Delta v}{4\delta} \sqrt{(2k\delta-1)^2-e^{-4k\delta}}.
\end{equation}
Here, $\delta$ is the interface width and depends on the fluid's specifics and the flow shear velocity. According to Eq.~(\ref{eq:RayleighFormula}), the instability only occurs for $k \delta \leq 0.64$, while the system is stable against perturbations with higher wavenumbers \cite{charru}. Similar to the PVM, Eq.~(\ref{eq:RayleighFormula}) recovers Kelvin's rate for $k\delta\ll1$.

\subsection{Point Vortex Model (PVM)}
We consider a two-dimensional superfluid containing N point vortices with quantized circulations $\Gamma=h/M$. When the inter-vortex separation is greater than a few healing lengths, vortices are advected by the velocity field created by other vortices. The equation of motion of each vortex is $d\vec{r}_i/dt = \vec{v_i}^0$, where $\vec{v_i}^0$ is the velocity field created by all the other vortices. If we consider a 1D array of equispaced vortices at coordinates $(x_n,y_n)=(d_v/2+n d_v, 0)$, moving in a 2D space of coordinates $(x,y)$ without boundaries, the tangential velocity of the superfluid flow can be written as: 
\begin{equation*}
    v_{x}(x, y) = -\frac{\Gamma}{2d_v} \frac{\sinh \left(2\pi y/d_v\right)}{\cosh \left(2\pi y/d_v\right)+\cos \left(2\pi x/d_v\right)}.
\end{equation*}
From this relation, the width of the shear layer is naturally expressed in units of $\delta=d_v/(2\pi)$, or equivalently $\delta=\hbar/(M\Delta v)$.

When considering the ring geometry, $\vec{v_i}^0$ must take into account the boundary conditions, namely that the flow must have a zero radial component at both the internal ($R_i$) and external ($R_e$) radii. We include the boundary conditions by using the method of image vortices \cite{supp}, and solve the equation of motions for the vortex necklace configuration in the ring with the Runge-Kutta method of fourth order. From the obtained trajectories of the vortices, $\vec{r}_i (t)$ we compute the normalized angular structure factor $s (m, t)$. 

Let us remark that vortices arranged in a ring array are not necessarily unstable, as noted by Havelock\cite{havelock}. When the vortex array encloses an inner boundary without circulation, the array is unstable only for $N\geq7$. Moreover, finite vortex number heavily suppresses the growth rate, converging to Eq. \ref{eq:pvm} for $N\gg 100$. On the other hand, when the array encloses the inner boundary having a circulation $w_i=N/2$, the array is unstable for $N\geq2$. In the latter setup, the growth rate converges to Eq. \ref{eq:pvm} already for $N\geq6$, justifying its application in the present work \cite{supp}. 

\subsection{Dissipative effects in the PVM}
The effect of dissipation on vortex dynamics can be introduced in the context of the two-fluid model as the effect induced by the mutual friction within the normal and superfluid components. This model, coined the dissipative PVM, includes two mutual friction coefficients $\alpha$ and $\alpha'$ associated with a dissipative term and a reactive term, respectively. Assuming the normal component is at rest, meaning, $v_n=0$, the dissipative PVM describes the motion of the vortices, $\frac{d \vec{r}_i}{dt} = (1-\alpha'\kappa_i^2) \vec{v}_i^0 - \alpha \kappa_i\hat z \times \vec{v}_i^0$. This modification introduces the correction factor $|\gamma_d|=\sqrt{(1-\alpha')^2+\alpha^2}$ to the growth rate given by Eq.~\eqref{eq:pvm} (see Supplementary Information). It is worth remarking that $|\gamma_d|$ can be either larger or smaller than 1, depending of the specific values of the mutual friction coefficients.

\subsection{Angular structure factor analysis}
At $t=0$, the one-dimensional angular structure factor of a finite array of $N_v$ vortices placed in a perfect necklace arrangement, with angular coordinates $\theta_j^0 = 2\pi j/N$, is $S^0(m)=\sin^2(\pi m)/(N_v\sin^2(\pi m /N_v))$. The departure from the necklace configuration can be modeled through the small fluctuations in the vortex positions at $t=0$: $\theta_j = \theta_j^0 + \delta \theta$. In crystals, small fluctuations ($\delta \theta \ll 2\pi/N_v$) are considered as a disorder of the first kind \cite{Warren}, and they modify the structure factor as $S(m,t) \approx S^d(m)-m^2\langle \delta \theta^2 \rangle(t)\,S^d(m)$, where $S^d(m)$ correspond to the structure factor of a given realization, and in general $S^d(m)\neq 0$, for different $m$. In the limit case, taking the average over many realizations, $S^d(m)\rightarrow S^0(m)$. Here, the temporal dependence of $S(m,t)$ is entirely provided by the term $\langle \delta \theta^2 \rangle(t)$. In the context of the PVM \cite{aref, havelock}, the motion of the vortices is linked to the the underlying shear-flow instability. In particular, the deviation from their initial position grows as $\delta \theta\sim e^{\sigma_m t}$, where $\sigma_m$ is given by Eq.~(\ref{eq:pvm}). Therefore, the temporal evolution of the structure factor is $S(m,t) \sim e^{2 \sigma_m t}$.

\vspace{-5pt}
\subsection{Maximum growth rate $\sigma^{*}$}
To obtain the maximum growth rate $\sigma^*$ experimentally, we fit the dispersion relation of the measured rates, Fig. \ref{fig:Fig2} c-e, using the following function $f(x, \sigma^*) = \sigma^* \frac{\sqrt{e^{-4\eta x}-(2\eta x-1)^2}}{A}$, with $x=m/\Delta w$, and $A=\max\left[\sqrt{e^{-4\eta x}-(2\eta x-1)^2}\right]=(W(e^{-1}) + 1)/(2\eta)\approx 0.639/\eta$, where $W(x)$ is the Lambert W-function and $\eta=0.8$ (see Supplementary Information\cite{supp} for details). The function $f(x,1)$ corresponds to Eq.~\eqref{eq:RayleighFormula} normalized to the maximum value shown as the magenta line in Fig.~\ref{fig:Fig2}c-e. We perform the fit of the dispersion relation letting $\sigma^*$ as the only free parameter.

\subsection{Lyapunov exponent}
To extract the Lyapunov exponent, $\Lambda$, of the system, we perform 40 PVM simulations under nearly identical conditions for a necklace with $N = 12$. The initial positions of each vortex are taken randomly within a range of one healing length ($0.5\,\mu$m) around their reference values for a perfectly periodic necklace. We then define the function $\mathcal{L}_k = \langle |\vec{r_i}^k-\vec{r_j}^k| \rangle_{i,j}$, as the average distance between two simulated trajectories of the $k$-th vortex in the necklace. Here, we denote by  $\left\langle \cdot \right \rangle_{i,j}$ the average over different simulations, i.e., $i,j=0,...,40$. Then, we compute the average over the $N$ vortices $\langle \Delta \vec{x} \rangle = \left\langle \mathcal{L}_k \right \rangle_{k}$, which we report in Fig.~\ref{fig:Fig4}d after normalizing it to the mean separation of any two points in the ring geometry $\mathrm{\bar{d}} = \int_{\Omega}\int_{\Omega}\sqrt{(x-x')^2+(y-y')^2} \,dx\,dy\,dx'dy' / A^2$, where $\Omega$ is the ring region with area $A=\pi (R_e^2-R_i^2)$. Although $\mathrm{\bar{d}}$ is straightforward to write, computing the integrals to obtain an analytical result is quite involved. For this reason, we numerically evaluate it by taking $10^5$ random points uniformly distributed inside the ring geometry delimited by $R_i$ and $R_e$. Then, we compute the $10^{10}$ possible combinations for the point-to-point distances and calculate their average value to estimate the mean separation $\mathrm{\bar{d}} \approx \SI{41.78\pm 0.02}{\mu m}$. We extract the characteristic rate $\Lambda$ from a fit of $\langle \Delta \vec{x}\, \rangle$ initial trend over the first \SI{14}{ms} from the starting time of the instability.

\section{Data availability}
The data that support the figures within this paper are available from the corresponding author upon reasonable request.

\section{Author contributions} 
D.H.R., G.D.P., F.S., F.M., and G.R. conceived the study. D.H.R., N.G., G.D.P., and W.J.K. performed the experiments. D.H.R. and N.G. analyzed the experimental data. D.H.R., K.X., C.F., and M.M. carried out numerical simulations. 
All authors contributed to the interpretation of the results and to the writing of the manuscript.

\section{Competing interests} 
The authors declare no competing interests.

%--------------------------------------------------
%------------------ Bibliography ------------------

%--------------------------------------------------
%------------------ External Figures --------------
%--------------------------------------------------
\renewcommand{\thefigure}{Ext.\arabic{figure}}
\setcounter{figure}{0}
\renewcommand{\theequation}{Ext.\arabic{equation}}
\setcounter{equation}{0}

\clearpage
\onecolumngrid

\begin{figure}[t!]
    \centering	
    \includegraphics[width=\textwidth]{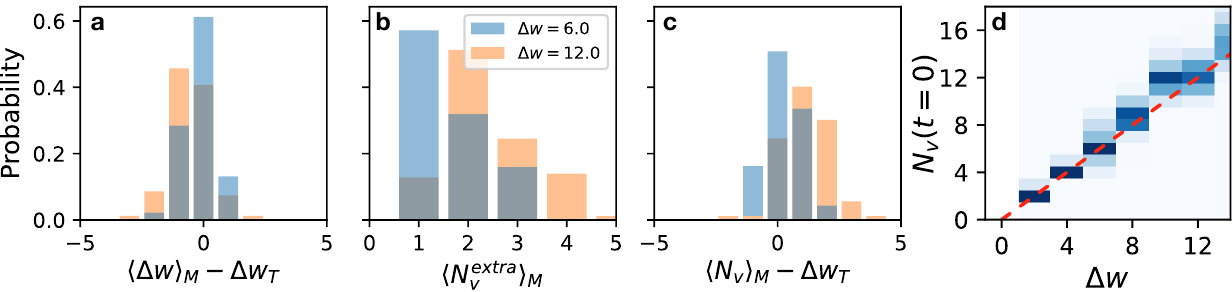}
        \caption{\textbf{a}, Infidelity in creating the target circulation state state, $\langle\Delta w\rangle _M - {\Delta w}_T$. \textbf{b}, Number of spurious vortices observed before removing the optical barrier, and \textbf{c}, Deviation of the total number of vortices from the target state, $\langle N_v \rangle _M - \Delta w_T$. All three panels were generated from 100 experimental repetitions for each of the two target states $\Delta w_T=6,12$ (blue and orange, respectively). \textbf{d}, Total number of vortices detected after removing the barrier, $t=0$ of vortex dynamics, as a function of the imprinted winding number difference $\Delta w_T$. The red dashed line is the identity line, $N_v=\Delta w$.
        }
\label{fig:ExpMeth2}
\end{figure}

\renewcommand{\thefigure}{S.\arabic{figure}}
\setcounter{figure}{0}
\renewcommand{\theequation}{S.\arabic{equation}}
\setcounter{equation}{0}
\renewcommand{\thesection}{S.\arabic{section}}
\setcounter{section}{0}
\renewcommand{\thetable}{S.\arabic{table}}
\setcounter{table}{0}

\setlength{\tabcolsep}{18pt}

\onecolumngrid
%\newpage

\setcounter{equation}{0}
\setcounter{figure}{0}
\setcounter{table}{0}

%%%%%%%%%%%%%%%%%%%%%%%%%%%%%%%%%%%%%%%%%%%%%%%%%%%%
\clearpage

\begin{center}
\textbf{\large Supplemental Information}
\end{center}
\normalsize 
\setcounter{page}{1}

\onecolumngrid
\section{Experimental Methods}

\paragraph{Thermodynamic properties}
The thermodynamic properties of the superfluid are obtained from analytical calculations based on the polytropic approximation for trapped gas, for which $\mu = g_{\gamma} n^{\gamma}$, where $\gamma$ is the effective polytropic index $\gamma = \partial \log \mu /\partial \log n$ \cite{Heiselberg2004,Haussmann2008}. In the BEC limit, $\gamma=1$, while at unitarity and in the BCS limit, $\gamma=2/3$. Within this approximation, the chemical potential and the Fermi energy take the form \cite{delpace2022imprinting}:
\begin{equation}\label{eq:chemicalpotential}
\mu_0  =  \left[\frac{ \Gamma(\frac{1}{\gamma} +\frac{3}{2})}{\pi^{3/2}\Gamma(\frac{1}{\gamma}+1)} \frac{\sqrt{M/2}\,\omega_z N g_{\gamma}^{1/\gamma}}{R_\mathrm{e}^2-R_\mathrm{i}^2}\right]^{\frac{2\gamma}{\gamma+2}},\quad E_F = 2\hbar\left[\frac{\hbar\omega_z N}{m(R_\mathrm{e}^2-R_\mathrm{i}^2)}\right]^{1/2}
\end{equation}
where $\Gamma$ is the Gamma function, $g_{\gamma}$ is a pre-factor that depends on the interaction regime: (i) for $(k_Fa)^{-1}>1$, $g_\mathrm{BEC} = 4\pi\hbar^2a_M/M$ with $a_M = 0.6 \, a$, $M = 2 m$, and $m$ is the mass of the ${}^6$Li atom; (ii) for $(k_Fa)^{-1}<-1$, $g_\mathrm{BCS} = \frac{\hbar^2}{2m}(6\pi^2)^{2/3}$; (iii) for $|(k_Fa)^{-1}|<1$, $g_\mathrm{UFG} = \xi_B \frac{\hbar^2}{2m}(6\pi^2)^{2/3}$, where $\xi_B$ is the Bertsch parameter taking the value $\xi_B \simeq 0.37$ at unitarity for $T=0$ \cite{Haussmann2008}. 
The Fermi wave number (see main text) is $k_F = \sqrt{2m\,E_F/\hbar^2}$.

\paragraph{Speed of sound}

\begin{figure}[h!]
	\centering	
 \vspace{10pt}
        \includegraphics[width=\textwidth]{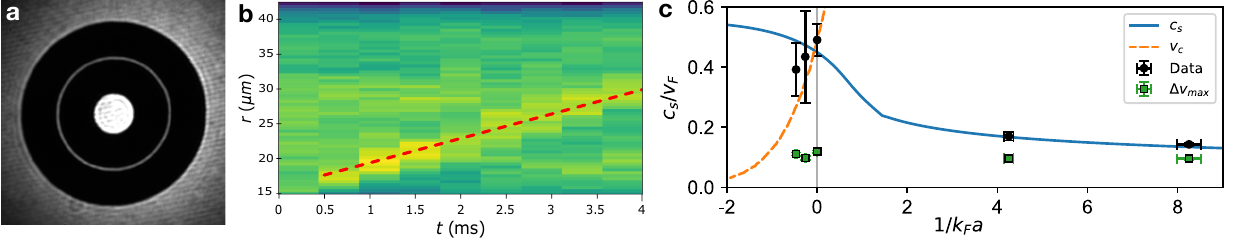}
            \caption{
            \textbf{a}, Image of the optical potential projected onto the atomic sample after being shaped by the DMD. \textbf{b}, Density wave propagating in the radial direction after abruptly enlarging and restoring the inner radius of the ring trap by $\SI{1.2(2)}{\mu m}$.
            \textbf{c}, Black circles correspond to the measured propagation velocity of the density wave as a function of the interaction parameter. Green square symbols correspond to the maximum relative velocity $\Delta v_\mathrm{max}$ at the interface explored in this work. The solid lines display the expected speed of sound $c_s/v_F$ in the BEC and crossover regimes (see text). The dashed line corresponds to the critical velocity for pair breaking in the BCS regime \cite{Weimer2015}. 
            }
	\label{fig:ExpMeth1}
\end{figure}

We measured the speed of sound by preparing the system in the same ring geometry of Fig. \ref{fig:ExpMeth1}a without the optical barrier separating the ring in two regions. We excite a sound wave propagating along the radial direction by abruptly enlarging and subsequently restoring the inner radius of the box potential. This procedure creates a density burst that travels at constant velocity along the radial direction toward the external radius. We measured the speed of sound $c_s$ by fitting the perturbed radial density profile with a Gaussian function, as shown in Fig. \ref{fig:ExpMeth1}b. The measured speed of propagation is reported in Fig. \ref{fig:ExpMeth1}c.
To calculate the speed of sound analytically, we calculate $c_s = \sqrt{\frac{n}{M}\frac{\partial \mu}{\partial n}} =\sqrt{\gamma \mu/M}$, and take the ratio with the Fermi velocity $v_F = \hbar k_F/m$.
In the BEC limit and for our geometry, the ratio $c_s/v_F$ can be expressed analytically in terms of $k_Fa$ and is given by $\frac{c_s}{v_F} = \left(\frac{3k_Fa_M}{2^{13/2}}\right)^{1/3}$, and for the BCS and crossover regimes as $\frac{c_s}{v_F}=\sqrt{{1 \over 3}\xi^{3/4}}$, where we assumed $\xi\equiv \xi(k_Fa)$ at $T=0$ \cite{Kwon2020}. The expected behavior of $\frac{c_s}{v_F}$ is shown in Fig. \ref{fig:ExpMeth1}c as a solid line.
For $1/k_Fa<0$, the measured speed is below the expected speed of sound. However, the measured propagation speed agrees with Leggett 3D homogeneous critical velocity for superfluid pair breaking \cite{Weimer2015}, suggesting this method fails to capture the speed of sound in the BCS regime. Nonetheless, the critical velocity for superfluid pair breaking is above the range of superfluid tangential velocities explored in this work.

%--------------------------------------------------
%--------------- Numerical Methods ----------------
%--------------------------------------------------

\section{Numerical Methods}
\subsection{Gross-Pitaevskii simulations}
In the BEC regime, at $T=0$, we simulate our system by solving the three-dimensional Gross-Pitaevskii (GP) equation
\begin{equation}
i\hbar \frac{\partial \psi(\bm{r},t)}{\partial t}= \left [  -\frac{\hbar^2}{2M}\nabla^2 + V(\bm{r},t) +g n(\bm{r},t) \right]\psi (\bm{r},t),
\label{eq:GPE}
\end{equation}
where $ V(\mathbf{r},t)$ is the external potential, $g n(\bm{r},t)$ is the mean-field potential due to the interaction strength $g=4\pi\hbar^2a_M/M$, the density is $n(\bm{r},t)=\int|\psi(\bm{r},t)|^2d^3\bm{r}$, and $\int n(\bm{r},t) d^3\bm{r}=N$ is the total number of molecules. In particular, the expression for the external potential we employed is: 

\begin{equation}\label{Vtrap}
V(r_\perp,z,t)=V_G(t) e^{-2\frac{(r_\perp-R_0)^2}{\sigma^2}} +V_r \left [ \tanh \left(\frac{r_\perp-R_{e}}{\epsilon} +1\right) +  \tanh \left(\frac{R_{i}-r_\perp}{\epsilon} +1\right)  \right] + 
\frac{1}{2} M (\omega^2_x x^2+\omega^2_y y^2+\omega^2_z z^2),
\end{equation}

where $r_\perp\equiv\sqrt{x^2 +y^2}$, $V_G(t)$ and $\sigma$ are the height and the size of the circular Gaussian barrier located at a distance $R_0=(R_{i}+R_{e})/2$ from the center in $xy$ plane, and $V_r$ and $\epsilon$ are the height and the stiffness of the ring potential in the $xy$ plane limited by internal and external radii $R_{i}$ and $R_{e}$ respectively. 
The simulation parameters are $a_M=1010\, a_0$, $N=3\times 10^4$. We used a grid of equal size along $x$ and $y$-directions equal to $\left[-60, 60\right]\mu$m and $ \left[-10,10\right]\mu$m along $z$ axis, based on up to $384 \times 384 \times 32$ grid points.
The harmonic potential frequencies are the same as in the experiment, $\omega_x=\omega_y=2\pi \times 2.5$~Hz and $\omega_z=2\pi \times 396$~Hz. The parameters of the ring potential are $V_r/h=2500$~Hz, $\epsilon=1.5$~$\mu$m, $R_{i}=10$~$\mu$m and $R_{e}=45$~$\mu$m respectively. 
The barrier size (FWHM) is $\sigma=1.4$~$\mu$m, whereas its height is linearly decreased in time during the barrier removal, starting from the initial value of $V_G(t=0)=V_0=h\times 5000$~Hz.

We first compute the ground-state of the condensate (with the circular barrier at $V_0$) by numerically minimizing the GP energy functional corresponding to Eq. (\ref{eq:GPE}) by means of a conjugate gradient algorithm \cite{Press2007,Modugno2003}. The ground-state chemical potential is $\mu/h=887$~Hz. Then, we imprint an opposite circulation in the two rings, adding a phase term $\exp( \pm i w \theta)=\exp[\pm i w  \arctan (y/x)]$ in the outer (+) and inner (-) ring, respectively, being $w$ the integer winding number. This phase imprinting results in an opposite velocity field in each of the two rings with $v=\pm\hbar/M\nabla \theta=\pm\hbar w/(Mr_\perp)$. 
The dynamical evolution is then triggered by linearly lowering the height $V_G(t)$ of the internal barrier at 28~ms, as done in the experiment. At the end of this phase, $N_v=\Delta w$ vortices are nucleated at $r_\perp=R_0$ forming a regular array with angular periodicity $\Delta \theta=2\pi/N_v$.

\subsection{Finite-temperature kinetic model}
At finite temperature, the system is partially condensed, and its wavefunction can be written as the sum of condensate and thermal components. Here, we use the collisionless Zaremba-Nikuni-Griffin model \cite{ZNGbook, JacksonPRA2002, NickJPhysB2008, nick_book} (cZNG) to investigate the effect of the thermal component. This model has already been successfully applied in the study of different phenomena such as the  collective modes \cite{JacksonPRL2002,JacksonPRL2001,XhaniPRL2020,XhaniPRR2022}, soliton and vortex dynamics \cite{JacksonPRA2007,JacksonPRA2009,AllenPRA2013,AllenPRA2014,XhaniPRL2020}. 
The condensate  wavefunction $\psi$  evolves according to the generalized Gross-Pitaevskii equation:
\begin{equation}
\label{gped}
i \hbar \frac{\partial \psi (\bm{r},t)}{\partial t}=\left[ - \frac{\hbar^2 \nabla ^2}{2M}+V(\bm{r},t)+g(n(\bm{r},t)+2n_\mathrm{th}(\bm{r},t))\right] \psi(\bm{r},t) \;,
\end{equation}
which includes an additional term with respect to the Eq.~\eqref{eq:GPE}: the mean-field potential of the thermal cloud ($2g n_{\mathrm th}$ where $n_\mathrm{th}$ is the thermal cloud density). The thermal cloud dynamics is instead described through the phase-space distribution function $f$, which satisfies the collisionless Boltzmann equation:
\begin{equation}
\label{bolt}
\frac{\partial f}{\partial t}+ \frac{\mathbf{p}}{M} \cdot \nabla_{\mathbf{r}} f-\nabla_{\mathbf{r}}V^{\rm eff} _{\rm th} \cdot \nabla_{\mathbf{p}}f=0
\end{equation}
where $V^{\rm eff}_{\rm th}=V+2g (n+n_{\rm th})$ is the generalized mean-field potential felt by the thermal particles. The thermal cloud density instead is found as $n_{\rm th}=1/(2\pi\hbar)^3 \int d\mathbf{p} ~ f(\mathbf{p},\mathbf{r},t)$). These two equations are solved self-consistently in a grid of equal size along the $x$ and $y$ directions, equal to  $\left[-106,106\right]\mu$m, and to $ \left[-22,22\right]\mu$m along the $z$ axis, based on $768 \times 768 \times 160$ grid points. The total particle number is equal to $N=3 \times 10^4$. 

We first find the condensate equilibrium density by solving the time-independent generalized GP equation:
\begin{equation}\label{gped2}
\mu \psi _0=\left( -\frac{\hbar ^2}{2M} \nabla ^2  + V + g (n_0 + 2 n^0_\mathrm{th})\right)  \psi_0,
\end{equation}
with  $n_0$ and $n_0^\mathrm{th}$ being the equilibrium condensate and thermal density, respectively, in the presence of the circular Gaussian barrier. The initial thermal cloud density Ansatz is based on a  Gaussian profile obtained for a certain temperature \cite{ZNGbook}. 
Then, we imprint a phase on the initial condensate wavefunction, as it was done for $T=0$ GP simulations, having  opposite signs in the outer (+) and inner (-) rings. After that, the generalized GP equation and the Boltzmann equation are solved self-consistently in order to describe the system dynamics, in the presence of a time-dependent circular barrier whose potential height is removed in 28\,ms. 
Due to the repulsive interaction between condensate and thermal particles, the thermal density $n^{\rm th}$ is maximum where the condensate density $n^{\rm BEC}$ is minimum, as it occurs e.g.~at the barrier position, at the edges of the condensate and at the vortex cores.

\subsection{Time evolution and stability analysis from the numerical results}

 \begin{figure}[ht!]
    \begin{center}
        \includegraphics[width=0.9\columnwidth]{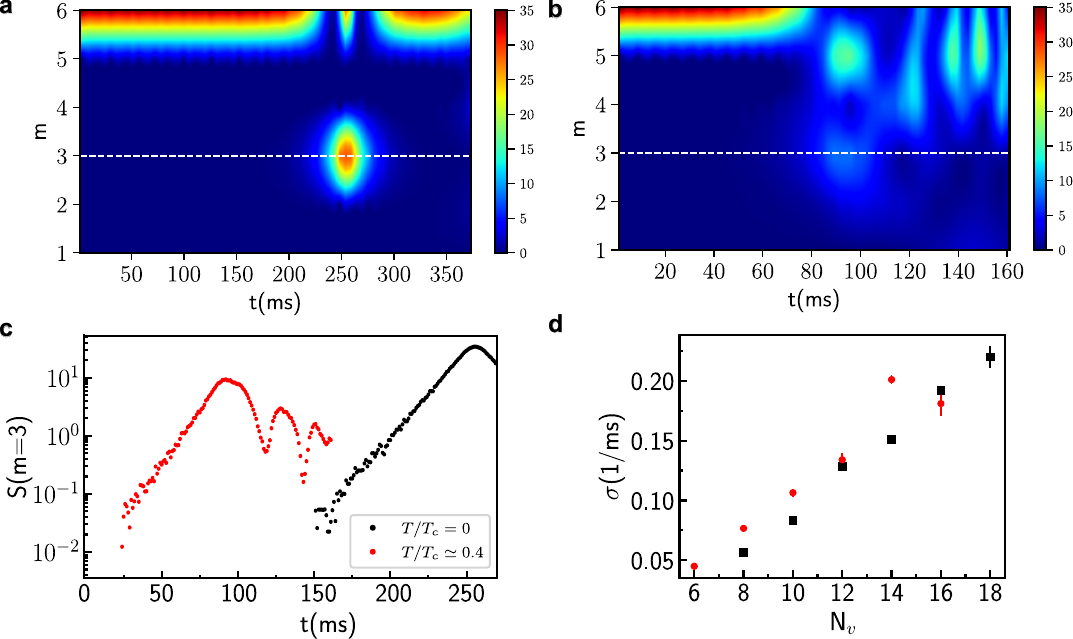}
            \caption{Heat map of the angular structure factor $S(m,t)$ for the $\Delta w=6$ configuration for: \textbf{a}, $T=0$ (GP) and \textbf{b}, $T\simeq 0.4 T_c$ (cZNG). The dashed line corresponds to $m=\Delta w/2$, whose amplitudes in logarithmic scales as a function of time for both temperatures are shown in panel \textbf{c}. \textbf{d}, Comparison between the growth rate extracted at $T=0$ and $T\simeq 0.4 T_c$ as a function of the number of vortices.}
        \label{fig:NumMeth3}
    \end{center}
\end{figure}

From the simulations performed either using the GP equations for the $T=0$ case, or the cZNG model for the $T\neq 0$ case, we extract the density profile integrated along $z$. Then we identify the positions of the vortices ad different evolution times $t$, from which we calculate the structure factor $S(m,t)$ for different modes $m$. The result is shown in Fig. \ref{fig:NumMeth3} a for the case of $\Delta w=6$ (indicated by the color). Fig.~\ref{fig:NumMeth3}c, shows the time-dependence of the structure factor for the mode  $m=\Delta w/2$ mode both for the $T=0$ (black) and $T=53nK$ (red) which is characterized by an initial exponential growth. Similarly, Fig. \ref{fig:NumMeth3} b shows the behaviour of the structure factor at temperature $T=53\, \mbox{nK} \simeq 0.4 \,  T_c$, close to the experimental value. We first observe that similar to the experimental results, at finite temperature, the growth of the $m=3$ mode amplitude starts almost immediately after the vortex array is generated, while at $T=0$ the instability starts much later. Furthermore, at a finite temperature, more modes are populated. 
We then extract the growth rate of the instability, and this procedure is repeated for different initial velocities, with the  results  shown in Fig. \ref{fig:NumMeth3}d. Interestingly, even though the starting time of the instability and the population of different modes are strongly affected by the presence of the thermal cloud, the growth rate of the $m=\Delta w/2$ mode turns out to be only slightly affected, being in average larger than the one at $T=0$. We also note that at longer time evolution, the presence of the thermal cloud significantly affects the vortices dynamics, consistent with the results of \cite{JacksonPRA2009, AllenPRA2013}.

Some additional tests have been performed to prove the stability of the GP results presented in this work. We verified that by doubling or halving the number of particles $N$ or the width of the internal barrier $\sigma$, the growth rates do not change. We also included in the GP simulations some noise in the density distribution of the ground state of the BEC. This noise term only affects the time when the instability starts without changing the growth rates of the most unstable mode. 
Moreover, cZNG simulations show that fixing the condensate number of the finite temperature case to be the same as the simulation at $T=0$ provides a similar growth rate, opposite to what is found in a Josephson junction \cite{XhaniPRR2022} where the superfluid dynamics strongly depends on the condensate number.

Finally, we tried to include in the simulation an \textit{imperfect} phase imprinting. This has been done by adding a phase to the ground state wavefunction distributed around a mean value $2\pi w$ with fluctuations smaller than $\pi$. This phase imprinting leads to the creation of density waves, which affect the dynamics of the vortex crystal. However, by averaging on different realizations, we found that the most unstable mode's growth rate has a value that is only $18\%$ smaller than the one of the \textit{clean} configuration. Nevertheless, we do not exclude that the presence of additional vortices due to the experimental imprinting techniques could have a larger effect on the observed growth rates.

\subsection{Thickness of the interface layer and correction factor}\label{sec:EffectiveWidth}

\begin{figure}[ht!]
    \begin{center}
        \includegraphics[width=0.7\columnwidth]{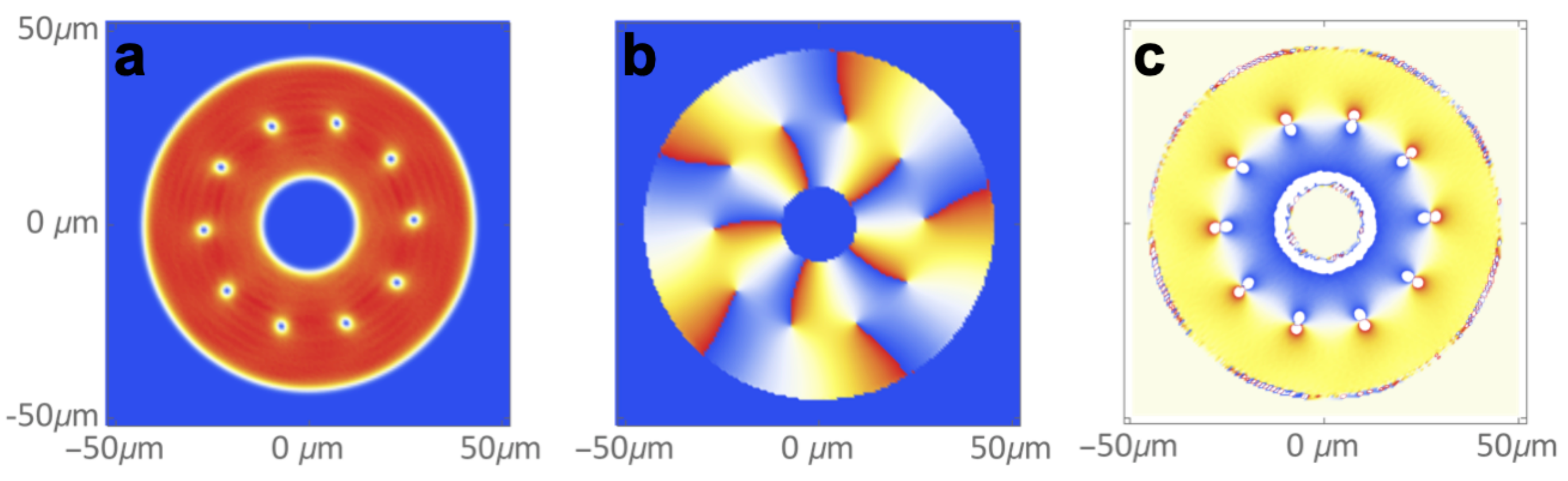}
            \caption{Gross-Pitaevskii simulations. \textbf{a}, Density distribution $n(x,y,0)$; \textbf{b}, Phase distribution $\phi(x,y,0)$; \textbf{c}, Angular velocity distribution $v_{\theta}(x,y,0)$ for the case $\Delta w=10$ at $t=0$.}
        \label{fig:NumMeth1}
    \end{center}
\end{figure}

\begin{figure}[h!]
    \begin{center}
    \vspace{10pt}
        \includegraphics[width=0.95\columnwidth]{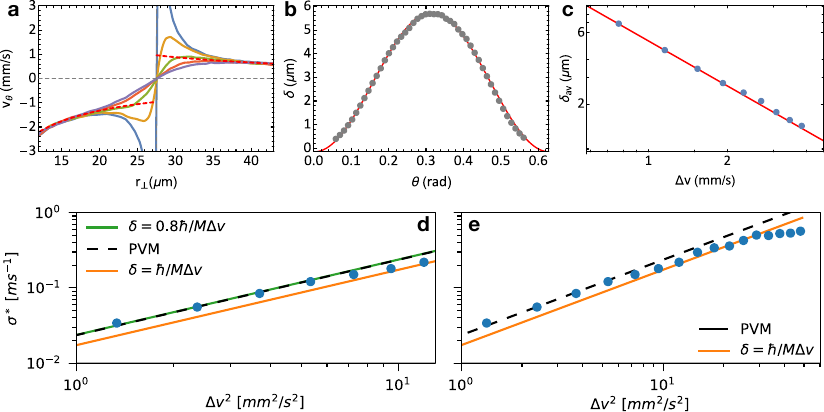}
            \caption{\textbf{a}, Plot of $v_{\theta}$ for different angular positions between two adjacent vortices for the case $N_v=10$. The vertical gray dashed line corresponds to $r=R_0$, and the dashed red line represents the imprinted velocity in the two rings with modulus $\hbar N_v /(2 M r_{\perp})$.
            \textbf{b}, Values of the interface half-width $\delta$ (gray data point) for different angular positions between two consecutive vortices located at $\theta=0$~rad and $\theta=0.63$~rad ($N_v$=10). This behaviour is fitted with a sinusoidal function $\delta= a \cos (N_v \theta)+b$ (red line) to extract the mean value $\delta_\text{av}=b$.
            \textbf{c}, Values of $\delta_\text{av}$ as a function of $\Delta v$, plotted together with the expected behavior $\delta=\hbar/(M \Delta v)$ (red line).
            \textbf{d--e}, Extracted growth rate from the GP simulations (blue circles) as a function of $\Delta v$ for the experimentally explored velocities (\textbf{d}), and all simulated velocities (\textbf{e}). The PVM maximum growth rate is shown as a black dashed line, while the solid lines correspond to Rayleigh's formula for the maximum growth rate with $\eta=0.8$ (green line) and $\eta=1$ (orange line).
            }
        \label{fig:NumMeth2}
    \end{center}
\end{figure}

To estimate the interface thickness $\delta$ in Eq.~(2), we analyzed the results of the GP simulations in the following way. From the phase $\phi$ of the condensate wavefunction, we numerically compute the velocity field as $v=\hbar/M \nabla \phi$. In Fig. \ref{fig:NumMeth1}, we show, from left to right, the density distribution $n(x,y,0)$, the phase $\phi(x,y,0)$, and the tangential velocity field $v_{\theta}(x,y,0)$, for the case with $\Delta w=10$ at $t=0$, namely immediately after the removal of the circular barrier. Considering that the tangential velocity field $v_{\theta}$ changes sign at $r_{\perp}=R_0$, to extract the width of the interface layer between the two regions, we fit the radial velocity profile with the function $v_{\theta}(r_{\perp})=a\tanh \left [ (r_{\perp}-R_0)/\delta(\theta) \right]/r_{\perp}$.
The behaviour of $v_{\theta}(r_{\perp})$ is shown in Fig. \ref{fig:NumMeth2}a for different values of $\theta$ between two adjacent vortices. 
The value of $\delta (\theta)$ extracted from the fit is shown in Fig. \ref{fig:NumMeth2}b. The interface thickness exhibits a sinusoidal behavior with minima in correspondence with the vortex cores and maxima in the middle of each vortex pair.

We define the effective value of the interface half-width $\delta_\text{av}$ as the average value of $\delta (\theta)$, and extract it by fitting such quantity with a sinusoidal function $\delta= a \cos (N_v \theta)+\delta_\text{av}$. 
In Fig.~\ref{fig:NumMeth2}c we report the obtained values of $\delta_\text{av}$ as a function of $\Delta v= \hbar \Delta w/ (MR_0)$. 
A similar study of the effective thickness of the interface can be performed within the PVM. As commented in the Methods, the equation of motion of a linear array of equispaced vortices can be analytically solved in a 2D space $x-y$ without boundaries, providing Eq. (3) of the main text for the velocity field. Plotting $v_x (x_0, y)$ for various $x_0 \in [-0.5 d_v, \, 0.5\, d_v]$ between two neighboring vortices provides similar trends of Fig.~\ref{fig:NumMeth2} a-b. We then employ the same procedure used to extract $\delta_\text{av}$ from the GP results for the PVM prediction, obtaining results in agreement with those of GP simulations. In particular, the estimated average interface thickness with both methods is well in agreement with the formula $\delta_\text{av} = \hbar /(M \Delta v)$, which we plot in Fig.~\ref{fig:NumMeth2}c as a red line.

On the other hand, Rayleigh's formula involves a linearly varying velocity between the two merging superfluids. We account for this difference by introducing an effective interface thickness $\delta = \eta \hbar /( M \Delta v)$, where $\eta$ is a phenomenological parameter of the order of $1$. We then fix the value of $\eta$ asking that the expressions of the most unstable mode under Rayleigh's formula [Eq.~(2)] and PVM are the same, namely:

\begin{equation}
    \sigma_R^*(k, \Delta v) =  \left(\frac{\sqrt{e^{-2\eta}-(\eta-1)^2}}{\eta}\right)\frac{M \Delta v^2}{4\hbar} = \sigma_\mathrm{PVM}(k^*, \Delta v)= \frac{1}{2} \frac{M\Delta v^2}{4\hbar}.
\label{Eq_SM:sigma}
\end{equation}
For the two expressions to be the same, we have to fix $\left(\sqrt{e^{-2\eta}-(\eta-1)^2}\right)/\eta = \frac{1}{2}$, yielding the value of $\eta \approx 0.80465$. 

Finally, we verified that such a value of $\delta$ is also consistent with GP simulation results of the most unstable mode $\sigma_\text{GP}$. In particular, in Fig.~\ref{fig:NumMeth2}e, we report the values of $\sigma_\text{GP}$ as extracted from GP simulations (symbols) and compare it with those given by Eq.~(\ref{Eq_SM:sigma}) with $\eta = 1$ (orange line) and $\eta = 0.8$ (green line). The latter is observed to well reproduce the GP results for low velocities, whereas for larger ones $\eta = 1$ matches the GP results better. The transition between these two regimes happens at velocities where the PVM is no longer applicable, i.e., when the inter-vortex distance is in the same order of magnitude as the healing length. The transition approximately occurs when $h/(M\Delta v) \approx 15~\xi$, with the associated velocity limit of $\Delta v= h/(15\,\xi M)\approx 3.16$~mm/s, or $\Delta w \approx 16$. For these reasons, we use the value $\eta \approx 0.8$ for $\Delta w < 16$ for all the measurements reported in this work.

\subsection{Point-Vortex Model (PVM) simulations}
\subsubsection{Model considering boundary conditions}

We consider a two-dimensional fluid containing $N_v$ point vortices with quantized circulations $\Gamma=h/M$. When the inter-vortex separation is greater than a few healing lengths, vortices are advected by the velocity field created by other vortices. The equation of motion of each vortex is $d\vec{r}_i/dt = \vec{v_i}^0$, where $\vec{v_i}^0$ is the velocity field created by all the other vortices. When considering the ring geometry, $\vec{v_i}^0$ must take into account the boundary conditions, namely that the flow must have a zero radial component at both the internal ($R_i$) and external ($R_e$) radii. We include the boundary conditions by using the method of image vortices \cite{Martikainen}:

\begin{align}
    \vec{v_i}^0 &= \frac{\Gamma}{2\pi} \sum_{i\neq j}^{N_v}  \hat{z} \times \frac{\vec{r}_i-\vec{r}_j}{\left|\vec{r}_i-\vec{r}_j\right|^2} +\frac{1}{2}\left(\frac{\Gamma}{2\pi} \sum_{j=1}^{N_v}  \hat{z} \times \frac{\vec{r}_i-\vec{0}}{\left|\vec{r}_i-\vec{0}\right|^2}\right)\\ \nonumber
    &+\frac{\Gamma}{2\pi} \sum_{j=1}^{N_v} \sum_{n=1}^{\infty} \hat{z} \times \left( \frac{\vec{r}_i-\left(\frac{R_e}{R_i}\right)^{2n}\vec{r}_j}{\left|\vec{r}_i-\left(\frac{R_e}{R_i}\right)^{2n}\vec{r}_j\right|^2}
    + \frac{\vec{r}_i-\left(\frac{R_i}{R_e}\right)^{2n}\vec{r}_j}{\left|\vec{r}_i-\left(\frac{R_i}{R_e}\right)^{2n}\vec{r}_j\right|^2}\right)\\ \nonumber
    &-\frac{\Gamma}{2\pi} \sum_{j=1}^{N_v} \sum_{n=0}^{\infty} \hat{z} \times \left(\frac{\vec{r}_i-\left(\frac{R_e}{R_i}\right)^{2n}\left(\frac{R_e^2}{|\vec{r}_j|^2}\right)\vec{r}_j}{\left|\vec{r}_i-\left(\frac{R_e}{R_i}\right)^{2n}\left(\frac{R_e^2}{|\vec{r}_j|^2}\right)\vec{r}_j\right|^2}
    +\frac{\vec{r}_i-\left(\frac{R_i}{R_e}\right)^{2n}\left(\frac{R_i^2}{|\vec{r}_j|^2}\right)\vec{r}_j}{\left|\vec{r}_i-\left(\frac{R_i}{R_e}\right)^{2n}\left(\frac{R_i^2}{|\vec{r}_j|^2}\right)\vec{r}_j\right|^2}\right)
\end{align}

\bigskip

We solve the equations of motions for the vortex necklace configuration in the ring with the Runge-Kutta method of fourth order. From the obtained trajectories of the vortices, $\vec{r}_i (t)$, we compute the normalized angular structure factor $s (m, t)$.

\subsubsection{Stability of the vortex array}\label{sec:Stability_PVM}

To study the stability of the vortex array and the subsequent expected growth rate, we perform a linear stability analysis around the array equilibrium position. 
As demonstrated by Havelock \cite{havelock}, the arrangement of the vortex polygon is a stationary configuration. Let us remark that the PVM can be rewritten in terms of complex variables \cite{aref} with the following mapping $z=x+iy$. This transforms the vortex equation of motion to be:
\begin{align}\label{eq:PVM_complex}
\frac{d z^*_a}{dt} &= \frac{\Gamma}{2\pi i} \left[\sum_{b\neq a}^N\frac{1}{z_a-z_b}+\frac{N}{2}\frac{1}{z_a}\right.\\
    &+\sum_{b=1}^N\sum_{n=1}^{\infty}\left(
        \frac{1}{z_a-\left(\frac{R_e}{R_i}\right)^{2n}z_b}
        +
        \frac{1}{z_a-\left(\frac{R_i}{R_e}\right)^{2n}z_b}\right)\\
    &\left.-\sum_{b=1}^N\sum_{n=0}^{\infty}\left(
    \frac{1}{z_a-\left(\frac{R_e}{R_i}\right)^{2n}\left(\frac{R_e^2}{|z_b|^2}\right)z_b}
    +
    \frac{1}{z_a-\left(\frac{R_i}{R_e}\right)^{2n}\left(\frac{R_i^2}{|z_b|^2}\right)z_b}
    \right)\right],
\end{align}
where $z^*$ denotes the complex conjugate of $z$. Setting the position of the $a$-th vortex to be $z_a^0(t) = R_0 e^{i\Omega t}e^{\frac{2\pi i a}{N}}$, where $\Omega$ naturally introduces the angular drift frequency the polygon array can undergo without breaking the array's symmetry. Substituting this Ansatz, we obtain the following equation for the angular frequency $\Omega$:
\begin{align}
\frac{8\pi \Omega R_0^2}{\Gamma} &= 2(N-1)+4\frac{N}{2}
- \frac{4N}{1-\left(\frac{R_i}{R_0}\right)^{2N}} - \frac{4N}{1-\left(\frac{R_e}{R_0}\right)^{2N}}\\
    &+4N\sum_{n=1}^{\infty}\left(
         \frac{1}{1-\left(\frac{R_i}{R_e}\right)^{2nN}}
        -\frac{1}{1-\left(\frac{R_i}{R_e}\right)^{2nN}\left(\frac{R_i}{R_0}\right)^{2N}}
        +\frac{1}{1-\left(\frac{R_e}{R_i}\right)^{2nN}} 
        -\frac{1}{1-\left(\frac{R_e}{R_1}\right)^{2nN}\left(\frac{R_e}{R_0}\right)^{2N}} 
        \right).
\end{align}\\
Let us remark on the following. The first term on the right-hand side corresponds to the contribution of the vortices inside the ring system. The second corresponds to the imaginary vortices located at the origin, fixing the circulation around the inner boundary to $w_i=N/2$, setting $w_i=0$ increases this term by a factor of 2, namely $4 N$. The third and fourth terms correspond to the contribution of the first imaginary vortex introduced in inner and outer boundaries, respectively. The fifth term corresponds to further corrections due to the extra imaginary vortices required to fulfill exactly the boundary conditions.

To analyze the stability of the polygon, let us write the perturbed positions as $z_a(t) = (1+\eta_a(t)) z_a^0(t)$. The equation for $\eta_a(t)$ is:
\begin{align}
\frac{d\eta_a^*}{dt} -i\Omega\eta_a^* &= \frac{i\Gamma}{2\pi R_0^2} z_a^0\left(\sum_{b\neq a}^N\frac{z_a^0\eta_a-z_b^0\eta_b}{(z_a^0-z_b^0)^2}+
\frac{N}{2}\frac{\eta_a}{z_a^0}\right.\\
    &+\sum_{b=1}^N\sum_{n=1}^{\infty}\left(
        \frac{z_a^0\eta_a-\rho_{ie}^nz_b^0\eta_b}{(z_a^0-\rho_{ie}^nz_b^0)^2} +
        \frac{z_a^0\eta_a-\rho_{ie}^{-n}z_b^0\eta_b}{(z_a^0-\rho_{ie}^{-n}z_b^0)^2} 
        \right)\\
    &\left.-\sum_{b=1}^N\sum_{n=0}^{\infty}\left(
        \frac{z_a^0\eta_a-\rho_{i0}\rho_{ie}^nz_b^0\eta_b}{(z_a^0-\rho_{i0}\rho_{ie}^nz_b^0)^2} +
        \frac{z_a^0\eta_a-\rho_{e0}\rho_{ie}^{-n}z_b^0\eta_b}{(z_a^0-\rho_{e0}\rho_{ie}^{-n}z_b^0)^2} 
    \right)\right)
\end{align}
where $\eta^*$ denotes the complex conjugate of $\eta$, and, for convenience, we defined $\rho_{ie} = (R_i/R_e)^2$, and $\rho_{i0}=(R_i/R_0)^2$, and $\rho_{e0}=(R_e/R_0)^2$. The latter equation can be mapped into:
\begin{equation}
\frac{d\vec{\eta}^*}{d t} -i\Omega\vec{\eta}^* = -i \textbf{C}\vec{\eta}    
\end{equation}
Here $\vec{\eta}$ is the vector whose $a$-th coordinate correspond to $\eta_a$, and $\textbf{C}$ is a matrix whose $(a,b)$ element is:
\begin{align}
    C_{a,b} &= A_{a,b} + \sum_{n=1}^{\infty} B_{n;a,b}\\
    A_{a,b} &= - 4\left(S_1+\frac{N}{2}-S_{4,0}-S_{5,0}\right) \delta_{ab} + 4(1-\delta_{ab})\frac{e^{2\pi i(b-a)/N}}{(1-e^{2\pi i(b-a)/N})^2}\\[10pt] \nonumber
    B_{n;a,b} &= 4\left( \frac{\rho_{ie}^ne^{2\pi i(b-a)/N}}{(1-\rho_{ie}^n e^{2\pi i(b-a)/N})^2} +\frac{\rho_{ie}^{-n} e^{2\pi i(b-a)/N} }{(1-\rho_{ie}^{-n} e^{2\pi i(b-a)/N})^2}\right.\\
    &\left.-\frac{\rho_{i0}\rho_{ie}^n e^{2\pi i(b-a)/N}}{(1-\rho_{i0}\rho_{ie}^n e^{2\pi i(b-a)/N})^2} -\frac{\rho_{e0}\rho_{ie}^{-n} e^{2\pi i(b-a)/N} }{(1-\rho_{e0}\rho_{ie}^{-n}e^{2\pi i(b-a)/N})^2} 
\right),
\end{align}
and the values $S_{i,n}$:
\begin{align}
\begin{split}
S_1 &= \sum_{\gamma=1}^{N-1}\frac{1}{(1-e^{2\pi i\gamma/N})^2}, \qquad \quad \; S_{2,n} = \sum_{\gamma=1}^{N} \frac{1}{(1-\rho_{ie}^n e^{2\pi i\gamma/N})^2},\\
S_{3,n} &= \sum_{\gamma=1}^{N} \frac{1}{(1-\rho_{ie}^{-n} e^{2\pi i\gamma/N})^2}, \qquad S_{4,n} = \sum_{\gamma=1}^{N} \frac{1}{(1-\rho_{i0}\rho_{ie}^n e^{2\pi i\gamma/N})^2},\\
S_{5,n} &= \sum_{\gamma=1}^{N} \frac{1}{(1-\rho_{e0}\rho_{ie}^n e^{2\pi i\gamma/N})^2}.
\end{split}
\end{align}

Considering $\vec{\eta}=\vec{\nu}+i\vec{\omega}$, we obtain:
\begin{equation}
\dot{\vec{\nu}}-i\dot{\vec{\omega}} =  i\left(\Omega \textbf{1} - \textbf{C}\right)\vec{\nu} +\left(\Omega\textbf{1}+\textbf{C}\right)\vec{\omega},
\end{equation}
where $\textbf{1}$ is the identity matrix. This can be mapped in the second-order equation:
\begin{equation}
\ddot{\vec{\nu}} = \left((\mathrm{Im}\textbf{C})^2 + (\mathrm{Re}\textbf{C})^2-\Omega^2\textbf{1}\right)\vec{\nu}
\end{equation}
Hence, setting $\vec{\nu}\sim e^{\lambda t}$, we get the eigenvalue equation:
\begin{equation}
\lambda^2 \vec{\nu} = \left((\mathrm{Im}\textbf{C})^2 + (\mathrm{Re}\textbf{C})^2-\Omega^2\textbf{1}\right)\vec{\nu}
\end{equation}
Efforts to analytically solve this eigenvalue problem in the cases of a single boundary, inner or outer, have been made by Havelock \cite{havelock}. However, this problem can be easily solved numerically, setting the following results for our values of $R_i=10\,\mu$m and $R_e=45\,\mu$m.
\begin{figure}[h!]\label{fig:grNg}
    \begin{center}
    \vspace{10pt}
        \includegraphics[width=0.95\columnwidth]{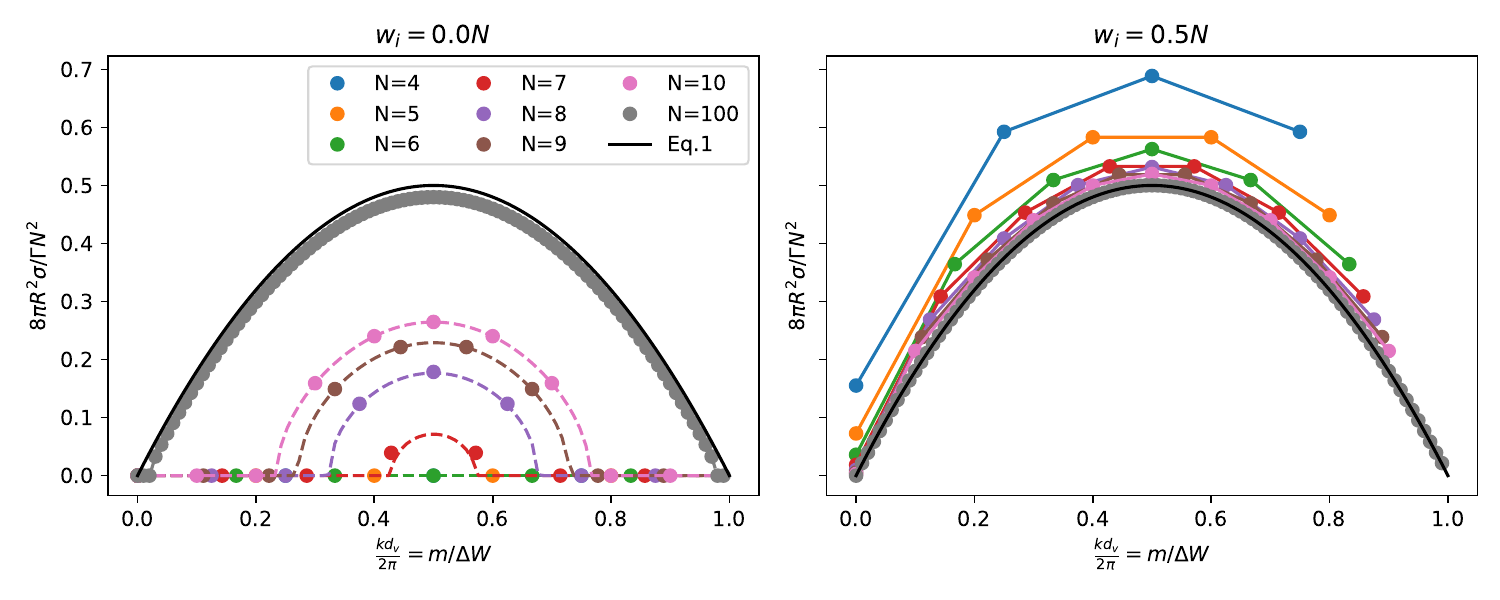}
            \caption{Instability growth rate including inner and outer boundaries calculated using PVM simulations (symbols). (Left) The case when the inner circulation is zero, $w_i=0$, compared with Havelock's analytical results \cite{havelock} (dashed lines). (Right) Experimental scenario with inner circulation $w_i=N/2$. The black line corresponds to Eq.~(1) of the manuscript.
            }
    \end{center}
\end{figure}

When the circulation around the inner boundary is zero, $w_i=0$, the first unstable mode happens for $N=7$ vortices, as shown in Fig.~S.5 (left panel). However, the flow created in the experiment has a circulation around the inner boundary with quantization $w_i=N/2$, modifying the growth rate of the instability as shown in Fig.~S.5 (right panel). Introducing this circulation makes the vortex array unstable even for $N=2$ vortices. Furthermore, for $N \approx 6$, the rates become similar to the limiting case of an infinite linear array of vortices provided by Eq.~(1) of the manuscript. In contrast, with $w_i=0$, order of 100 vortices are required to approach such a limit (Fig.~S.5, left panel).

\subsubsection{Dissipative effects in the PVM}
Dissipative vortex dynamics can be accounted for in the context of the two-fluid model by considering the effect of the mutual friction between the normal and superfluid components. The PVM can be extended into the so-called dissipative PVM, containing two mutual friction coefficients $\alpha$ and $\alpha'$, associated to a dissipative term and a reactive term, respectively. The equation of motion prescribed by the dissipative PVM is:

\begin{equation}
    \frac{d \vec{r}_i}{dt} = (1-\alpha') \vec{v}_i^0 - \alpha \hat z \times \vec{v}_i^0.
\end{equation}
This equation can be rewritten in the complex form, $z=x+iy$, as follows:
\begin{equation}
    \frac{d z_a^*}{dt} = (1-\alpha'+i \alpha) \vec{v}_a^* = \gamma_d \vec{v}_a^*,
\end{equation}
where $\gamma_d = 1-\alpha'+i \alpha$, and $\vec{v}_a^*$ corresponds to the right-hand side of (S.9-S.11) in the scenario given by the ring geometry. Following the stability analysis from the previous section we obtain the analog of Eqs.~(S.19) and (S.23-24):
\begin{equation}
\frac{d\vec{\eta}^*}{d t} -i\Omega\vec{\eta}^* = -i \gamma_d\textbf{C}\vec{\eta}.    
\end{equation}
Hence:
\begin{equation}
\ddot{\vec{\nu}} = \left((\mathrm{Im}\gamma_d\textbf{C})^2 + (\mathrm{Re}\gamma_d\textbf{C})^2-\Omega^2\textbf{1}\right)\vec{\nu} = |\gamma_d|^2\left((\mathrm{Im}\textbf{C})^2 + (\mathrm{Re}\textbf{C})^2-\frac{\Omega^2}{|\gamma_d|^2}\textbf{1}\right)\vec{\nu}
\end{equation}
Modifying the eigenvalue equation as follows:
\begin{equation}
\left(\frac{\lambda}{|\gamma_d|}\right)^2 \vec{\nu} = \left((\mathrm{Im}\textbf{C})^2 + (\mathrm{Re}\textbf{C})^2-\left(\frac{\Omega}{|\gamma_d|}\right)^2\textbf{1}\right)\vec{\nu}
\end{equation}

This modifications to the eigenvalue equation correct the eigenvalues with the factor $|\gamma_d|=\sqrt{(1-\alpha')^2+\alpha^2}$. Thus, the growth rate of the instability for the dissipative PVM is well approximated by:
\begin{equation}
\sigma_k = \sqrt{(1-\alpha')^2+\alpha^2}\,\sigma_k^0
\end{equation}
with $\sigma_k^0$ the growth rate for the ideal PVM, as shown in Fig~S.6. It is important to note that $|\gamma_d|$ can be either larger or smaller than 1 depending of the specific values of the mutual friction coefficients.

%--------------------------------------------------
%------------------ Bibliography ------------------
%--------------------------------------------------


\end{document}